\documentclass{jpp}
\usepackage{graphicx}
\usepackage{epstopdf, epsfig}
\usepackage{amsmath,amsfonts,amssymb}
\usepackage{hyperref}
\usepackage{amsmath,amsfonts,amssymb}
\usepackage[dvipsnames]{xcolor}
\usepackage[flushmargin,hang]{footmisc}
\usepackage{float}
\usepackage[font = default, labelfont = default]{caption}
\usepackage{subcaption}
\usepackage{bigfoot}
\usepackage{bm}

\usepackage[utf8]{inputenc}
\usepackage[T1]{fontenc}
\usepackage{amsmath}

\usepackage{tikz}
\usepackage{tikz-3dplot}
\usetikzlibrary{calc,patterns,angles,quotes}
\usetikzlibrary{decorations.pathreplacing,calligraphy}
\usetikzlibrary{decorations.pathreplacing}
\usetikzlibrary{decorations.markings}
\usetikzlibrary{decorations.pathmorphing}
\usetikzlibrary{3d,calc}
\usetikzlibrary{patterns,hobby,arrows.meta}
\usetikzlibrary{shapes.misc}
\usetikzlibrary{intersections}

\hypersetup{
    colorlinks=false,        
    pdfborder={0 0 1},       
    citebordercolor={0 1 0}, 
    linkbordercolor={1 0 0}, 
    urlbordercolor={0 0 1}   
}

\renewcommand{\footnotesize}{\scriptsize}

\newcommand{\im}{\text{i}}
\newcommand{\boldhat}[1]{\hat{\boldsymbol{#1}}}
\newcommand{\bsym}[1]{\boldsymbol{#1}}
\newcommand{\vths}[1]{v_{\text{th}, #1}}
\newcommand{\rmd}{{\rm d}}

\makeatletter
\renewcommand*{\@fnsymbol}[1]{%
  \ifcase#1
  \or \dagger 
  \or \ddagger 
  \or \S 
  \or \P 
  \or \| 
  \or ** 
  \or \dagger\dagger 
  \or \ddagger\ddagger 
  \else\@ctrerr
  \fi}
\makeatother

\shorttitle{Gyrokinetic Theory of Linear Gravitational Flute Interchanges}

\shortauthor{Z. Y. Tan and I. G. Abel}

\title{
    Gyrokinetic Theory of Linear Gravitational Flute Interchanges with Flow Shear Stabilisation 
}

\author{
    Z. Y. Tan\aff{1}\textsuperscript{,}\aff{2}\corresp{\email{zytan@umd.edu}},
    \and I. G. Abel\aff{2}\textsuperscript{,}\aff{3}
  }

\affiliation{

\aff{1}
Department of Physics, University of Maryland, College Park, Maryland 20742, USA

\aff{2}
Institute for Research in Electronics \& Applied Physics,
University of Maryland,
College Park, MD 20740, USA

\aff{3}
Gridfire Inc,
3422 Old Capitol Trail, Suite 700,
Wilmington, DE 19808, USA

}

\begin{document}

\maketitle

\begin{abstract}
    A collisionless electrostatic gyrokinetic theory is developed to describe how the presence of a differential velocity shear can help stabilise linear gravitational flute interchanges in slab geometry. This is made possible because the velocity shear acts to increase the perpendicular wavenumber of the unstable modes over time. Eventually, a threshold wavenumber is crossed, where the effect of gyroaveraging, captured by the Bessel functions appearing in the gyrokinetic equation, results in a damping of the instability by the nature of these Bessel functions behaving as decaying sinusoids. However, transient amplification, responsible for subcritical turbulence, can still occur. Numerical comparisons are made with a magnetohydrodynamic model with gyroviscous corrections as well as the GX gyrokinetic code. It is demonstrated that an increasing shear acts not only to accelerate the stabilisation effect but also to reduce the overall transient amplification.
\end{abstract}

\renewcommand{\thefootnote}{\textsuperscript{\arabic{footnote}}}

\section{Introduction}

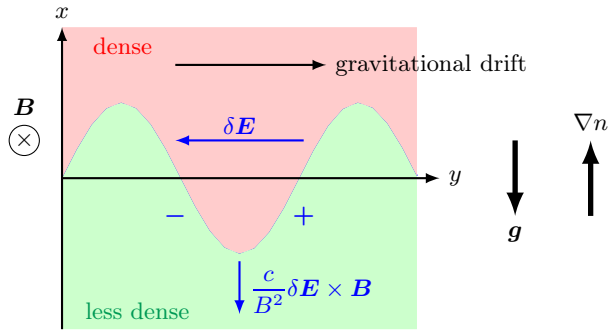
\begin{figure}
    \centering
    \begin{tikzpicture}
        \draw[thick, blue, samples=200, domain=0:4.7] plot (\x,{sin(2* \x r) + 2});
        \fill[red!20, opacity=0.3] 
        (0,4) -- plot[domain=0:4.7] (\x,{sin(2* \x r) + 2}) -- (4.7,4) -- cycle;
        
        \fill[green!20, opacity=0.3] 
        (0,0) -- plot[domain=0:4.7] (\x,{sin(2* \x r) + 2}) -- (4.7,0) -- cycle;
        
        \node[red] at (0.8,3.75) {dense};
        \node[ForestGreen] at (1,0.25) {less dense};
        
        \draw[thick, -latex] (0, 2) -- (5, 2) node[anchor = west]{\(y\)};
        \draw[thick, -latex] (0, 0) -- (0, 4) node[anchor = south]{\(x\)};

        \draw (-0.5, 2.5) circle(0.2);
        \node at (-0.5, 2.5) {\(\times\)};
        \node at (-0.5, 2.95) {\(\boldsymbol{B}\)};
        
        \draw[line width = 2pt, -latex] (6, 2.5) -- (6, 1.5) node[anchor = north]{\(\boldsymbol{g}\)};
        
        \draw[line width = 2pt, -latex] (7, 1.5) -- (7, 2.5) node[anchor = south]{\(\nabla n\)};
        
        \draw[thick, -latex] (1.5, 3.5) -- (3.5, 3.5) node[anchor = west]{gravitational drift};
        
        \node[blue] at (1.5, 1.5) {\(\boldsymbol{-}\)};
        \node[blue] at (3.2, 1.5) {\(\boldsymbol{+}\)};
        
        \draw[thick, -latex, blue] (3.2, 2.5) -- (1.5, 2.5);
        \node[blue] at (2.35, 2.7) {\(\delta\boldsymbol{E}\)};
        
        \draw[thick, -latex, blue] (2.35, 0.9) -- (2.35, 0.2);
        \node[blue] at (3.3, 0.5){\(\displaystyle \frac{c}{B^2} \delta\boldsymbol{E}\times\boldsymbol{B}\)};
    \end{tikzpicture}
    \captionsetup{width = 0.9\textwidth}
    \caption{Schematic of plasma setup in slab geometry for the onset of the gravitational flute interchange instability. \(\boldsymbol{B}\) is the mean magnetic field (into the page), \(\boldsymbol{g} = -g\hat{\boldsymbol{x}}\) is the gravitational field, and \(\nabla n\) is the density gradient. A wavelike perturbation (in blue) in the \(xy\) plane breaks the translational symmetry in \(y\), leading to a charge build up (indicated by the `\(\boldsymbol{-}\)' and `\(\boldsymbol{+}\)' symbols) due to the gravitational drift in which an opposing perturbed electric field \(\delta\boldsymbol{E}\) must be set up to maintain quasineutrality; however, the resulting \(c\delta\boldsymbol{E}\times\boldsymbol{B}/B^2\) drift amplifies the perturbation leading to an instability. The density gradient is necessary to ensure that \(\delta\bm{E}\) is pointing in the right direction. If the density gradient was opposite to that in the diagram, then direction of \(\delta\bm{E}\) would be flipped, resulting in a restoring rather than destabilising drift.}
    \label{fig1}
\end{figure}

The gravitational interchange instability was first analysed by \citet{Kruskal53} using ideal magnetohydrodynamics (MHD). This instability arises in configurations where the mean density gradient \(\nabla n\) of the plasma is in the opposite direction to the gravitational field \(\boldsymbol{g} = - g\hat{\boldsymbol{x}}\) as depicted in \autoref{fig1}. In this setup, we consider a quasineutral plasma in slab geometry, with a constant magnetic field \(\boldsymbol{B} = B\hat{\boldsymbol{b}} = B\hat{\boldsymbol{z}}\) threading it such that \(\boldsymbol{g}\perp \boldsymbol{B}\). Since part of the Larmor orbit is against the gravitational field, this gives rise to a `gravitational drift' \(\boldsymbol{g}\times \hat{\boldsymbol{b}}/\Omega_s = g\hat{\boldsymbol{y}}/\Omega_s\), where \(\Omega_s\) is the gyrofrequency of a particle of species \(s\)  \citep{Goldston95}. Because \(m_e \ll m_i\), this drift affects the ions more than the electrons, leading to charge separation when there is a wave-like perturbation in the presence of a density gradient. As depicted in \autoref{fig1}, the resulting perturbed electric field \(\delta\boldsymbol{E}\) and its corresponding drift amplifies the perturbation, causing an instability. This instability is purely electrostatic and does not perturb the magnetic field.

\par 
The above is an intuitive mental picture in the MHD context that allows us to visualise the instability. Unfortunately, it cannot be the full picture in gyrokinetics where quasineutrality is enforced as a constraint (rather than as an approximation from Gauss's law in MHD); hence, no charge separation can ever exist \citet{Abel13}. To understand the instability in this context, let us first consider the case in which the electrons have an adiabatic response. In this case, the electron density perturbation \(\delta n_e\) is proportional to the perturbed electric potential \(\delta \varphi\). As a result, an ion density perturbation \(\delta n_i\) would trigger an immediate electron perturbation \(\delta n_e\), resulting in a \(\delta\varphi\) (increasing from left to right in \autoref{fig1}) and producing the perturbed electric field \(\delta\bm{E}\). The situation is more opaque for flute modes since the electrons no longer have an adiabatic response and are now purely kinetic. As a result, unlike the direct proportionality relation \(\delta n_e \propto \delta\varphi\) previously, it is now not explicitly clear from the linear electron gyrokinetic equation what the relation between \(\delta\varphi\) and \(\delta n_e\) is. Nevertheless, the instability must manifest in a similar way; otherwise, there would be deviations from the MHD dispersion relation.\footnote{Since the plasma must remain quasineutral, if we are unable to exploit the adiabatic response of the electrons, the alternative must be an opposing drift that counteracts the gravitational drift. We think that a suitable candidate for this would be the polarisation drift. However, this is more difficult to visualise and slightly more unintuitive.}

\par
The first attempt to include finite Larmor radius (FLR) effects into the theory of this instability was performed by \citet{Rosenbluth62} using Vlasov theory. Their result predicted a \(\mathcal{O}(k_\perp \rho_i)\) correction to the original dispersion relation of \citet{Kruskal53}, where \(k_\perp\) is the perpendicular wavenumber and \(\rho_i\) is the thermal ion Larmor radius. While the ideal MHD result predicted no stable modes, the correction resulted in a threshold value of \(k_\perp \rho_i\), beyond which the modes were rendered stable. This is known as `FLR stabilisation' and it was reproduced by \cite{RobTay62} by modifying the stress tensor in the ideal MHD momentum equation. This modification is known as the \textit{gyroviscous stress}. It was first identified by \citet{Thompson61} and more rigorously treated by \citet{Braginskii65} in the context of collisional plasmas. Since then, a great deal of research has gone into modelling and understanding FLR stabilisation for the gravitationally driven interchange \citep{Rosenbluth65, Bowers68, Ariel69, Huba96, Schnack06, Ferraro06, Ferraro07, Xu16}.

\par 
In the context of fusion devices, gravitational effects are generally negligible; however, from our description in the beginning, we see that the instability may be induced by any drift that affects ions and electrons differently, such as the curvature or \(\nabla B\) drifts. The presence of such interchanges in these devices makes them interesting to study \citep{Rosenbluth57, Wakatani92, Zagorski99, Tatsuno99, Levitt04, Poli08}. Of particular interest is the  centrifugal mirror, where plasma rotation is imposed in order to confine the plasma \citep{Lehnert71, Hassam97, Ellis01, Teodorescu08}. The idea is that in the co-rotating frame, particles experience a centrifugal force. The component of this force perpendicular to the mirror field lines is balanced by the \(\boldsymbol{J \times B}\) force that manifests from the plasma rotation. On the other hand, the parallel component is unbalanced, with its direction pointing towards the centre of the device, thereby confining the plasma.

\par 
Unfortunately, because the electron centrifugal drift is negligible due to its small mass, it is mainly the ions that are affected \citep{Goldston95, Levitt04, Abel13}. Consequently, this centrifugal drift can drive the interchange instability. However, the presence of rotational shear can induce FLR stabilisation \citep{Hassam92, Ellis01, Huang01} by tilting the unstable modes and pushing \(k_\perp \rho_i\) past the aforementioned threshold. This results in the overall stability depending on the competition between the destabilising centrifugal drive and the stabilising shear. 

\par 
The idea of utilising sheared flows to suppress instabilities and turbulence has been researched thoroughly throughout the years \citep{Biglari90, ArtunTang92, Waltz98, Hobbs01, Newton10, Ivanov25}. However, even if shear stabilisation ensures that all modes are eventually stable, it is still possible for a stable mode to experience an initial positive growth before decaying. This is known as linear transient amplification and it is a possible driver of a phenomenon known as `subcritical turbulence' i.e. turbulence in the absence of linear instabilities \citep{Highcock10, Highcock11, Barnes11, Alex12, Landreman15}. Study of this feature is pivotal for the current problem, since some time is needed before the shear pushes the modes into the FLR stabilisation regime, during which transient amplification can occur. This was demonstrated by \citet{Ng05} and will become apparent in the numerical solutions in this work.

\par 
However, because the model by \citet{Ng05} only provides \(\mathcal{O}(k_y\rho_i)\) corrections to the MHD model by \citet{Kruskal53} (which takes \(k_y\rho_i \to 0\)), it is natural to question its validity as \(k_\perp\rho_i \) will inevitably approach \(\mathcal{O}(1)\) in the presence of shear. This is the theoretical motivation for this paper, which is to model the interchange using gyrokinetics \citep{AntonsenLane80, FriemenChen82, Catto81, Abel13}, which naturally provides \(k_y\rho_i\) corrections to all orders. This linear flow shear problem, in the framework of gyrokinetics, can then be analysed using the theory developed by \citet{Alex12} for ion temperature gradient and parallel velocity gradient instabilities.

\par 
We shall organize the paper as follows. In \S\ref{SECT: Assume}, we will describe the setup and assumptions made in the paper. In \S\ref{SECT: MHD gyro}, we will demonstrate how the \(\mathcal{O}(k_y\rho_i)\) correction in gyroviscous MHD leads to FLR stabilisation for the interchange in the absence of shear. Also, we will derive a time-dependent model to describe the problem with flow shear. These are then repeated in \S\ref{SECT: GK}, but this time using gyrokinetics. We will then compare the solutions from MHD and gyrokinetics with the GX gyrokinetic code \cite{Noah18, Noah24} in \S\ref{SECT: numerical}. Finally, we will conclude with a summary and a discussion on future work in \S\ref{SECT: conclusion}. We note that, throughout this paper, our notation will adhere strictly to \citet{Abel13} and \citet{Alex12} as far as possible. Any deviations will be highlighted in the paper.


\section{Assumptions and Setup} \label{SECT: Assume}

We draw the following assumptions from the previous studies on the interchange instability mentioned in the introduction. Firstly, we will consider a quasineutral hydrogen plasma \(Z_i = + 1, Z_e = -1, n_e = n_i\) where \(Z_i\) is the ionic charge, \(Z_e\) is the electron charge, and \(n_s\) is the mean density of species \(s\). We assume that the variation of \(n_s\) is purely in the \(\hat{\boldsymbol{x}}\) direction i.e. \(n_s = n_s(x)\), and that no temperature gradient exists so that the focus is solely on the interaction between the density gradient and the gravitational field. The plasma will be also assumed to be thermally equilibrated i.e. \(T_i = T_e\) where \(T_i\) and \(T_e\) are the ion and electron temperatures respectively.

\par 
Secondly, as depicted in \autoref{fig1}, the plasma is subjected to a uniform mean magnetic field \(\boldsymbol{B} = B\hat{\boldsymbol{b}} = B\hat{\boldsymbol{z}}\) and a constant gravitational field \(\boldsymbol{g} = - g \hat{\boldsymbol{x}}\) in which \(g \sim \vths{i}^2/a_N\) \citep{Ng05}. Here, \(\vths{s} = \sqrt{2T_s/m_s}\) is the thermal velocity of a particle of species \(s\) and \(a_N\) is the characteristic length scale of the plasma. In addition, we will consider a linear shear \(\boldsymbol{u} = Sx \boldhat{y}\), with \(S\) being the shear parameter such that \(\bm{u}\cdot\nabla \approx S x \partial / \partial y\) \citep{Alex12}. 

\par 
Thirdly, as explained in the introduction, we consider purely electrostatic fluctuations i.e. \(\delta\boldsymbol{B} = 0\). Moreover, we will only consider the most unstable interchanges (flutes), which are fundamentally 2D perturbations with \(k_\parallel \to 0\) \citep{Rosenbluth65, Hassam92}, where \(k_\parallel\) is the parallel wavenumber. Consequently, we need to consider full kinetic electrons since the parallel streaming needed for an adiabatic response is now not possible.

\par 
Fourthly, we shall focus solely on the linear collisionless theory as was done in earlier references \citep{Rosenbluth62, RobTay62, Ng05}. It is worth noting that the `collision-free' approximation of the Braginskii stress tensor is simply the aforementioned gyroviscosity \citep{RobTay62, Schnack06, parra19a}. Also, even though MHD fundamentally requires collisions, the collisionless assumption in our theory is still valid for two reasons: first, the primary difference between a collisional and collisionless model is the presence of Landau damping \citep{Xu16}, which our problem does not contain because we have taken \(k_\parallel \to 0\).\footnote{Note that we have taken the ion mean free path \(\lambda_{\text{mfp}i}\) to be large enough such that \(k_\parallel \lambda_{\text{mfp}i}\gg 1\) in order to maintain the collisionless property of the plasma.}; second, the perpendicular dynamics of the plasma is still fluid-like even in the absence of collisions, provided that \(\omega \ll \Omega_i\) and \(k_\perp \rho_i \ll 1\) \citep{Thompson61, Alex09}.

\par 
Lastly, we shall work with the low Mach number limit of gyrokinetics \citep{Abel13} such that we can independently control the shear and the gravitational drive. As we shall soon see, if we do not consider this limit, the `shearing box transformation' performed in \S\ref{MHD SECT: shear box} will not be able to cleanly remove the shear \(S\) from our equations due to the presence of centrifugal effects in the gyrokinetic equation, resulting in additional complexities.


\section{The MHD interchange with gyroviscosity} \label{SECT: MHD gyro}

The goal of this section is to understand how, in the absence of flow shear, FLR stabilisation arises when gyroviscosity is included in ideal MHD and how it introduces a threshold perpendicular wavenumber, beyond which the modes become stable. In addition, a finite-flow-shear time-dependent MHD model will also be derived. This model will be later used in \S\ref{SECT: numerical} to demonstrate how a mode with an initially unstable perpendicular wavenumber can be made stable with flow shear. This happens because the flow shear will eventually push the wavenumber past the aforementioned threshold.

\par 
The finite-flow-shear model in this section is similar to that of \citet{Ng05}; however, we have elected to re-derive the time-dependent equations from the MHD equations used in \citet{RobTay62} which, unlike \citet{Ng05}, does not assume incompressibility of the plasma. The reason is that it will be later demonstrated that, at zero shear, it is our model that produces the same dispersion relation as in gyrokinetics. As gyrokinetics is the focus of our paper, we shall leave the full derivation of our MHD model to \autoref{appx MHD} and only quote results from it.

\subsection{FLR stabilisation in the absence of flow shear} \label{SECT: MHD gyro - zero shear}

The original calculation of the interchange modes by \citet{Kruskal53} using the linearized ideal MHD equations. These modes were demonstrated to be unstable modes with the growth rate being:
\begin{equation}
    \begin{aligned}
        \gamma_g 
        =
        \sqrt{\frac{ \omega_{*i} \omega_{gi}}{k_y^2 \rho_i^2/2}}
        =
        \sqrt{g \frac{\rmd \log n_i}{\rmd x}}
        ,
    \end{aligned}
    \label{MHD:eqn1}
\end{equation}
where we have defined the frequencies
\begin{align}
    &\omega_{*s} = \frac{v_\text{th,s}}{2} k_y \rho_s \frac{ \rmd \log n_s}{ \rmd x}
    ,
    \label{MHD:eqn2}
    \\
    \intertext{and}
    &\omega_{gs} = \frac{g}{v_\text{th,s}} k_y \rho_s
    .
    \label{MHD:eqn3}
\end{align}
Later calculations by \citet{Rosenbluth62} using Vlasov theory derived a \(\mathcal{O}(k_y \rho_i)\) correction to the ideal MHD dispersion relation:
\begin{equation}
    \begin{aligned}
        \omega^2 
        =
        -
    	\frac{\omega_{gi} \omega_{*i}}{k_y^2 \rho_i^2/2}
        +
        \underbrace{(\omega_{gi} + \omega_{*i}) \omega}_{\mathcal{O}(k_y \rho_i)}
    	+
        \mathcal{O}(k_y^2 \rho_i^2)
        ,
    \end{aligned}
    \label{MHD:eqn4}
\end{equation}
this result was later reproduced by \citet{RobTay62}, using MHD but modifying the momentum equation to include gyroviscous stress \citep{Braginskii65, parra19a}. 

\par 
The derivation of \eqref{MHD:eqn4} can be found in \S\ref{appx MHD SECT: zero shear}. In this derivation, we have repeated the calculation in \citet{RobTay62}, but allowed for \(k_x \neq 0\). What follows is that the denominator of the first term on the RHS of \eqref{MHD:eqn4} is replaced by \(k_\perp^2 \rho_i^2/2\). This implies that having \(k_x \neq 0\) contributes to the stability of the mode. Our focus will be on the most unstable modes in which \(k_x = 0\) and hence \(k_\perp = k_y\).

\par 
We also note here that the \(\mathcal{O}(k_y \rho_i)\) correction in \eqref{MHD:eqn4} differs from equation (13) in \citet{Ng05} in which the latter does not possess the \(\omega_{gi}\omega\) term. As demonstrated in \autoref{appx Ng MHD}, this difference arises from the incompressibility assumption made in \citet{Ng05}. 

\par 
The dispersion relation \eqref{MHD:eqn4} can be solved to give
\begin{equation}
    \begin{aligned}
        &
        \omega 
        =
        \frac{1}{2} \left[
            \left(
            \omega_{gi} + \omega_{*i} 
            \right)
            \pm 
            \sqrt{
                \left(
            \omega_{gi} + \omega_{*i}
            \right)^2
                -
                4 \frac{\omega_{gi} \omega_{*i}}{k_y^2 \rho_i^2/2}
            }
        \right]
        .
    \end{aligned}
    \label{MHD:eqn5}
\end{equation}
We see that the system is stable when
\begin{equation}
    \begin{aligned}
        &
        (\omega_{gi} + \omega_{*i})^2 
        \geq
         \frac{8 \omega_{gi} \omega_{*i}}{k_y^2 \rho_i^2}
        \Longrightarrow
        \lvert k_y \rho_i \rvert
        \geq
         \frac{\sqrt{8 \omega_{gi} \omega_{*i}}}{\omega_{gi} + \omega_{*i}}
         .
    \end{aligned}
    \label{MHD:eqn6}
\end{equation}
Equation \eqref{MHD:eqn6} defines the threshold value of \(k_y \rho_i\) (equivalently \(k_\perp\rho_i\)) for FLR stabilisation. This was absent in \eqref{MHD:eqn1} and arises due to the addition of gyroviscosity to the MHD momentum equation, which accounts for first order FLR corrections (see \S\ref{appx MHD SECT: zero shear}).

\par 
In the next section, we outline a time-dependent theory in MHD with finite flow shear which predicts that an initially unstable mode with \(k_x = 0\) will evolve to have a finite \(k_x\) and eventually become stable when \(k_\perp \rho_i\) satisfies \eqref{MHD:eqn6}.

\subsection{Finite flow shear}

\subsubsection{Shearing boxes and time-dependent wavenumbers} \label{MHD SECT: shear box}

Let us first demonstrate how flow shear causes the wavenumbers \(k_\perp \rho_i\) to increase. We introduce the `shearing box transformation' \citep{Alex12, Fox17}, which corresponds to a coordinate transformation \((t, \mathbf{r}) \rightarrow (t', \mathbf{r}')\) to a local orthogonal Cartesian frame moving with and in the vicinity of some flux surface.\footnote{In our case, the `flux surfaces' are planes labelled by \(x\).} As mentioned in \S\ref{SECT: Assume}, we are considering a linear shear \(\boldsymbol{u} = S x \boldhat{y}\) and the transformation is given by:
\begin{equation}
    \begin{aligned}
		t' = t, \qquad 
        x' = x, \qquad 
        y' = y - Sxt, \qquad
        z' = z.
	\end{aligned}
    \label{MHD:eqn7}
\end{equation}
As seen in \S\ref{appx MHD SECT: time dep}, this transformation removes the \(Sx \partial / \partial y\) terms of the linearised MHD equations which would have hindered Fourier transforming it. This allows us to utilise the normal mode solutions as was done in the zero shear scenario. However, as the problem is now time-dependent, we do not assume a normal mode time-dependence, since the growth rates will also be time-dependent. Instead, any perturbative quantity assumes the form:
\begin{equation}
    \begin{aligned}
		X(\mathbf{r}', t')
        = 
        \hat{X}(t') e^{\im \boldsymbol{k}_\perp' \cdot \boldsymbol{r}'}
        =
        \hat{X}(t') e^{\im \boldsymbol{k}_\perp(\boldsymbol{k}'_\perp, t') \cdot \boldsymbol{r}}
        ,
	\end{aligned}
    \label{MHD:eqn8}
\end{equation}
where we remind the reader that we have taken \(k_\parallel = k_z \rightarrow 0\) for flute modes.

\par 
From the respective partial derivatives, we see that \(k_y = k_y'\), and crucially:
\begin{equation}
    \begin{aligned}
		k_x = k_x(t) = k_x' - St' k_y'
        .
	\end{aligned}
    \label{MHD:eqn9}
\end{equation}
This demonstrates the time-dependent property of \(k_x\) due to flow shear. As we explained in the zero shear case, we can set \(k_x' = 0\) to focus on the most unstable modes. However, even though we will initially have \(k_x = k_x' = 0\), \(\lvert k_x \rvert\) will increase over time as \(S \neq 0\) according to \eqref{MHD:eqn9}, meaning that \(k_x\) will not remain zero.

\par
This immediately poses a challenge for the gyroviscous MHD model which assumes \(k_\perp \rho_i \ll 1\) and only provides a \(\mathcal{O}(k_\perp \rho_i)\) correction. As \(t\) gets large, it is clear from \eqref{MHD:eqn9} that we will eventually reach \(k_\perp \rho_i \sim 1\) at which point our MHD theory breaks down. This is the motivation for a gyrokinetic treatment which does not suffer the same predicament.

\subsubsection{Time-dependent equation}

The original derivation of the finite-flow-shear time-dependent MHD equation was calculated by \citet{Ng05}. However, as mentioned in \S\ref{SECT: MHD gyro - zero shear}, because of the incompressibility assumption made in their paper, the \(\mathcal{O}(k_y \rho_i)\) correction to the zero shear dispersion relation differs from that of \citet{Rosenbluth62} and \citet{RobTay62} which is given by \eqref{MHD:eqn4}. More importantly, we will later see in \S\ref{GK SECT - zero shear}, that it is \eqref{MHD:eqn4} which agrees with our derivation of the analogous zero shear dispersion relation in gyrokinetics up to \(\mathcal{O}(k_y \rho_i)\).

For the above reasons, we have chosen to re-derive the time-dependent equation using the extended MHD equations in \citet{RobTay62} that replaces the incompressibility condition with the generalized Ohm's law. Substituting the solutions \eqref{MHD:eqn8} into these equations, solving for the perturbed ion density \(\widehat{\delta n}_i\), and suppressing all the primes gives us
\begin{equation}
    \begin{aligned}[b]
        &
        \frac{\partial^2 }{\partial t^2} \frac{\widehat{\delta n}_i(t)}{\widehat{\delta n}_i(0)}
        +
        \Bigg[
            \frac{2 S^2 t }{S^2 t^2 + 1}
            - 
            \im (\omega_{gi} + \omega_{*i})
        \Bigg]\frac{\partial}{\partial t} \frac{\widehat{\delta n}_i(t)}{\widehat{\delta n}_i(0)}
        \\
        &
        -
        \frac{1}{S^2 t^2 + 1}
        \Bigg[
        	\im \left(2\omega_{gi} + \omega_{*i}\right) S^2 t
            +
        	  \frac{\omega_{gi} \omega_{*i}}{k_y^2\rho_i^2/2} 
        \Bigg] \frac{\widehat{\delta n}_i(t)}{\widehat{\delta n}_i(0)}
        =
        0
        .
    \end{aligned}
    \label{MHD:eqn10}
\end{equation}
The derivation of this equation can be found in \S\ref{appx MHD SECT: time dep}. Though we have derived an \(\mathcal{O}(k_y^2 \rho_i^2)\) equation in \eqref{appx MHD: eqn33}, we have elected to use only the \(\mathcal{O}(k_y \rho_i)\) equations for reasons that will be explained in \S\ref{GK SECT - zero shear}.

\subsubsection{Short-time limit} \label{MHD SECT: short time} 

Similar to \citet{Alex12}, we would like to derive a short-time asymptotic limit in our model. A sensible way to think of what constitutes this is to consider the fact that the shear effects are weak; hence, the time-dependent lab frame \(k_x\) should exhibit weak time-dependence. Looking at \eqref{MHD:eqn9}, this would be equivalent to taking the limit \(St \ll 1\). By taking this limit and assuming solutions of the form \(e^{\im k_y y - \im \omega t}\), it can be shown that the zero-shear dispersion relation \eqref{MHD:eqn4} is recovered. Outside the short-time regime, we need to solve equation \eqref{MHD:eqn10} numerically. However, this will be deferred to \S\ref{SECT: numerical} as we want to first introduce our gyrokinetic model, which is the main objective of this paper.

\section{Gyrokinetic theory of gravitational flute interchanges} \label{SECT: GK}

In the previous section, we used MHD with gyroviscosity to obtain a \(\mathcal{O}(k_y \rho_i)\) correction to the ideal MHD dispersion relation. We saw that in the case of finite flow shear \(S \neq 0\), \(k_\perp \rho_i\) becomes time-dependent and eventually increases past a point where \(k_\perp \rho_i \sim 1\) in which the \(\mathcal{O}(k_y \rho_i)\) correction no longer suffices. To account for this shortfall, we shall perform an analogous calculation in gyrokinetics, which retains all orders of \(k_y\rho_i\).

\subsection{Linear collisionless gyrokinetic equation with gravity}

The inclusion of gravity results in an additional drift term \(\boldsymbol{g}\cdot\partial f_s/\partial\boldsymbol{v}\) on the LHS of the usual Fokker-Planck kinetic equation. Following the systematic procedure in \citet{Abel13}, we may then convert the equation to rotating gyrokinetic variables: the guiding-centre position \(\boldsymbol{R}_s\), particle energy \(\varepsilon_s\), magnetic moment \(\mu_s\), and the gyrophase \(\vartheta\). Next, we perform a perturbative expansion of the Fokker-Planck equation to \(\mathcal{O}(\epsilon^2 \Omega_s f_s)\) where \(\epsilon = \rho_i/a_N\) is the gyrokinetic expansion parameter with \(a_N\) being the characteristic length scale of the system. Finally, considering only the fluctuating part of the equation, ignoring collisions, and neglecting products of fluctuating quantities, we obtain the following linear collisionless gyrokinetic equation in the \textit{low Mach number} limit\footnote{The flux surface label and the symmetry direction are defined to be \(\psi = Bax\) and \(\nabla\phi = \hat{\boldsymbol{y}}/a\) respectively in order to draw parallels with the toroidal decomposition of the magnetic field (\(\boldsymbol{B} = \nabla\psi\times\nabla\phi\)) found in \citet{Abel13}.}:
\begin{equation}
    \begin{aligned}
		&
        \left[\frac{\partial}{\partial t} + \boldsymbol{u}(\boldsymbol{R}_s) \cdot \frac{\partial}{\partial 	   \boldsymbol{R}_s} \right] h_s
		+
		\left( w_\parallel \hat{\boldsymbol{z}} + \frac{g}{\Omega_s} \hat{\boldsymbol{y}} \right)\cdot \frac{\partial h_s}{\partial \boldsymbol{R}_s} 
		\\
		&=
		\frac{Z_s e F_{0s}}{T_s}\left[
		\frac{\partial }{\partial t} 
		+
		\boldsymbol{u}(\boldsymbol{R}_s) \cdot 
		\frac{\partial }{\partial \boldsymbol{R}_s}
		\right]\left\langle  \delta\varphi  \right\rangle_{\boldsymbol{R}}
		+\frac{c F_{0s}}{B}
        \left(
        \frac{\rmd \log n_s}{\rmd x} + \frac{m_s g}{T_s}
        \right) 
        \boldhat{y} \cdot 
        \frac{\partial \langle \delta\varphi \rangle_{\boldsymbol{R}}}{\partial \boldsymbol{R}_s}
        \, .
	\end{aligned}
    \label{GK:eqn1}
\end{equation}
This equation is analogous to equation (248) in \citet{Abel13}, where the terms multiplied by \(g\) account for the gravitational effects. The last term on the LHS accounts for the gravitational drift of guiding-centres while the last term on the RHS arises due to an additional \(m_s g x\) term added to \(\varepsilon_s\). In equation \eqref{GK:eqn1}, \(\langle \cdot \rangle_{\boldsymbol{R}}\) indicates a gyroaveraged quantity, \(\delta\varphi\) is the fluctuating electrostatic potential, \(w_\parallel\) is the parallel \textit{peculiar velocity} of the particle\footnote{The peculiar velocity \(\boldsymbol{w}\) is related to the actual particle velocity \(\boldsymbol{v}\) via \(\boldsymbol{w} = \boldsymbol{v} - \boldsymbol{u}\).}, and \(F_{0s}\) is the equilibrium Maxwellian distribution given by:
\begin{equation}
    F_{0s} (x(\boldsymbol{R}_s), \varepsilon_s) 
    =
    \frac{n_s(x(\boldsymbol{R}_s))}{\pi^{3/2}v_{\text{th},s}^3}
    e^{m_s g x/T_s} e^{-m_s w^2/ 2 T_s}
    .
    \label{GK:eqn2}
\end{equation}
The quantity \(h_s\) represents the gyrophase-independent distribution of Larmor rings of species \(s\), which is related to the overall distribution function \(f_s\) by:
\begin{equation}
    \begin{aligned}
        f_s = F_{0s} + \delta \! f_s = 
        F_{0s} - \frac{Z_s e \delta\varphi}{T_s} F_{0s} + h_s
        .
    \end{aligned}
    \label{GK:eqn2.1}
\end{equation}
It is worthwhile to note that, unlike in \citet{Abel13}, we do not possess the \(\mathcal{O}(\epsilon)\) correction to the mean electric field \(\boldsymbol{E}\), which they denote as \(\varphi_0\). This correction arises due to centrifugal and Coriolis effects, both of which are absent in our problem due to the simplified slab geometry.

\par 
The gyrokinetic equation \eqref{GK:eqn1} contains two unknowns \(h_s\) and \(\langle \delta\varphi \rangle_{\boldsymbol{R}}\). To close the system, we require the quasineutrality condition:
\begin{equation}
    \sum_s \frac{Z_s^2 e^2 n_s}{T_s} \delta \varphi
		=
		\sum_s Z_s e \int \rmd^3\boldsymbol{w} \,\left\langle h_s \right\rangle_{\boldsymbol{r} }
        .
    \label{GK:eqn3}
\end{equation}
Analogous to our earlier analysis in MHD, we shall first consider the case of zero shear and then move on to consider finite shear.

\subsection{Zero flow shear} \label{GK SECT - zero shear}

First, we consider the scenario where \(\boldsymbol{u} = 0\). In this case, \eqref{GK:eqn1} becomes
\begin{equation}
    \begin{aligned}
		&
        \frac{\partial h_s}{\partial t}  
		+
		\left( w_\parallel \hat{\boldsymbol{z}} + \frac{g}{\Omega_s} \hat{\boldsymbol{y}} \right)\cdot \frac{\partial h_s}{\partial \boldsymbol{R}_s} 
		=
		\frac{Z_s e F_{0s}}{T_s} 
		\frac{\partial \left\langle  \delta\varphi  \right\rangle_{\boldsymbol{R}}}{\partial t} 
		+
        \frac{c F_{0s}}{B}
        \left(
        \frac{\rmd \log n_s}{\rmd x} + \frac{m_s g}{T_s}
        \right)
        \boldhat{y} \cdot 
        \frac{\partial \langle \delta\varphi \rangle_{\boldsymbol{R}}}{\partial \boldsymbol{R}_s}
        .
	\end{aligned}
    \label{GK:eqn4}
\end{equation}
As done in MHD, we seek solutions of the form \eqref{appx MHD: eqn18}, setting \(k_x = 0\) for the most unstable modes as previously discussed:
\begin{align}
    &
    \delta\varphi (\boldsymbol{r},t)
    = 
    \widehat{\delta\varphi} \, e^{\im k_y y - \im\omega t}
    ,
    \label{GK:eqn5}
    \\
    &
    h_s (\boldsymbol{R}_s, t)
    = 
    \hat{h}_s e^{\im k_y \boldsymbol{R}_s\cdot\boldhat{y} - \im\omega t}
    .
    \label{GK:eqn6}
\end{align}
Our flux surface label in this problem can be defined as \(\psi = \psi(x) = Bax\). Because we intend to later perform the `shearing box transformation' discussed in \S\ref{MHD SECT: shear box} for the case of finite shear (which is an expansion in the vicinity of some flux surface), we perform a local approximation about \(x = 0\) \citep{Rosenbluth62} by considering \(x\sim\rho_i\). As a result, we can drop the \(m_s g x \sim \epsilon T_s\) term in the exponent of the Maxwellian in \eqref{GK:eqn2} and simply have \(F_{0s} = n_s e^{-w^2/\vths{s}^2} / (\pi\vths{s}^2)^{3/2} \). 

\par 
The gyroaverages of \(\delta\varphi\) and \(h_s\) are then computed in the usual way:
\begin{align}
    &
    \left\langle \delta\varphi \right\rangle_{\boldsymbol{R}}
        = 
    \widehat{\delta\varphi} \, e^{\im k_y \boldsymbol{R}_s\cdot\boldhat{y} - \im\omega t}
    J_0 \left(\frac{w_\perp k_y}{\Omega_s}\right)
    ,
    \label{GK:eqn7}
    \\
    &
    \left\langle h_s \right\rangle_{\boldsymbol{r}}
        = 
        \hat{h}_s e^{\im k_y y  - \im\omega t}J_0 \left(\frac{w_\perp k_y}{\Omega_s}\right)
    ,
    \label{GK:eqn8}
\end{align}
where we note that \(h_s\) is gyroaveraged at fixed \(\boldsymbol{r}\) while \(\delta\varphi\) is gyroaveraged at fixed \(\boldsymbol{R}_s\). 
Substituting \eqref{GK:eqn6} and \eqref{GK:eqn7} into the gyrokinetic equation \eqref{GK:eqn4} gives us an equation for \(\hat{h}_s\):
\begin{equation}
    \begin{aligned}
        \hat{h}_s
        &=
        \frac{Z_s e}{T_s} \left(
		\frac{
				\omega 
                - 
                \omega_{*s}
				-
				\omega_{gs}
		}{
			\omega  - \omega_{gs}
		}
		\right)
		J_0 \left(\frac{w_\perp k_y}{\Omega_s} \right)
        F_{0s}
		\widehat{\delta\varphi}
        ,
    \end{aligned}
    \label{GK:eqn9}
\end{equation}
where \(\omega_{gs}\) and \(\omega_{*i}\) were previously defined in \eqref{MHD:eqn2} and \eqref{MHD:eqn3} respectively. 

\par 
Substituting \eqref{GK:eqn5}, \eqref{GK:eqn8}, and \eqref{GK:eqn9} into the quasineutrality condition \eqref{GK:eqn3} gives us
\begin{equation}
    \begin{aligned}
        &
        \sum_s \frac{Z_s^2 e^2 n_s}{T_s} 
		=
		\sum_s 
    	\frac{Z_s^2 e^2}{T_s} 
    	\frac{n_s}{\pi^{3/2} v_{\text{th}, s}^3} 
    	\left(
    	\frac{
    		\omega 
    		- 
    		\omega_{*s}
    		-
    		\omega_{gs}
    	}{
    		\omega  - \omega_{gs}
    	}
    	\right)
    	\int \rmd^3\boldsymbol{w} \, 
    	\left[ J_0 \left(\frac{w_\perp k_y}{\Omega_s} \right) \right]^2
    	e^{-w^2/v_{\text{th}, s}^2}
        .
    \end{aligned}
    \label{GK:eqn10}
\end{equation}
The \(w_\parallel\) integral is simply a Gaussian integral while the \(w_\perp\) integral can be evaluated by utilising the following identity \citep{AbramowitzStegun}:
\begin{equation}
    \begin{aligned}
        \int_0^\infty \rmd x \left[
            x J_n(px) J_n(qx) e^{-a^2 x^2} 
        \right]
        =
        \frac{1}{2a^2} I_n \left(\frac{pq}{2a^2}\right) \exp\left(
            -\frac{p^2 + q^2}{4a^2}
        \right)
        ,
    \end{aligned}
    \label{GK:eqn11}
\end{equation}
where \(J_n\) and \(I_n\) are the Bessel functions and modified Bessel functions of the first kind, respectively. This allows us to obtain the following dispersion relation
\begin{equation}
    \begin{aligned}
        &
        \sum_s \frac{Z_s^2 e^2 n_s}{T_s} 
		=
		\sum_s 
        \frac{Z_s^2 e^2 n_s}{T_s}
        \frac{\omega - \omega_{gs} - \omega_{*s}}{\omega - \omega_{gs}} 
        I_0 \left(\frac{k_y^2 \rho_s^2}{2}\right)
        e^{-k_y^2 \rho_s^2/2}
        .
    \end{aligned}
    \label{GK:eqn12}
\end{equation}

\subsubsection{Long wavelength and small mass ratio limit}

We will now demonstrate how the zero shear dispersion relation in MHD from \S\ref{SECT: MHD gyro - zero shear} can be recovered from \eqref{GK:eqn12}. For a hydrogen plasma, taking the long wavelength limit i.e. \(k_y \rho_i \ll 1\) and hence \(I_0(k_y^2 \rho_i^2/2)  e^{-k_y^2 \rho_s^2/2}\approx 1 - k_y^2 \rho_i^2/2\) results in \eqref{GK:eqn12} becoming:
\begin{equation}
    \begin{aligned}
        &
        \frac{1}{T_i} + \frac{1}{T_e}
		=
        \frac{1}{T_i}
        \frac{\omega - \omega_{gi} - \omega_{*i}}{\omega - \omega_{gi}} 
         \left(1 - \frac{k_y^2 \rho_i^2}{2}\right)
        +
        \frac{1}{T_e}
        \frac{\omega - \omega_{ge} - \omega_{*e}}{\omega - \omega_{ge}} 
         \left(1 - \frac{k_y^2 \rho_e^2}{2}\right)
         .
    \end{aligned}
    \label{GK:eqn13}
\end{equation}
Taking \(m_e/m_i \ll 1\) and noting that
\begin{align}
    &
    \omega_{ge} = - \omega_{gi} \frac{m_e}{m_i}
    ,
    \label{GK:eqn14}
    \\
    &
    \omega_{*e} = - \omega_{*i} \frac{T_e}{T_i}
    ,
    \label{GK:eqn15}
    \\
    &
    \rho_e = -\rho_i \sqrt{\frac{m_e}{m_i} \frac{T_e}{T_i}}
    ,
    \label{GK:eqn16}
\end{align}
equation \eqref{GK:eqn13} becomes:
\begin{equation}
    \begin{aligned}
        &
        1 + \frac{T_i}{T_e}
		=
        \frac{\omega - \omega_{gi} - \omega_{*i}}{\omega - \omega_{gi}} 
         \left(1 - \frac{k_y^2 \rho_i^2}{2}\right)
        +
        \frac{T_i}{T_e}
        \frac{\omega + \frac{T_e}{T_i}\omega_{*i}}{\omega } 
        .
    \end{aligned}
    \label{GK:eqn17}
\end{equation}
Setting \(T_i = T_e\), the result is
\begin{equation}
    \begin{aligned}
        &
        \omega^2  = -\frac{\omega_{*i}\omega_{gi}}{k_y^2 \rho_i^2/2} + (\omega_{gi} + \omega_{*i})\omega
        +
        \mathcal{O}(k_y^2 \rho_i^2)
        ,
    \end{aligned}
    \label{GK:eqn18}
\end{equation}
which is exactly the MHD result in \eqref{MHD:eqn4}.

\par 
We note here that the gyrokinetic dispersion relation \eqref{GK:eqn17} and the MHD dispersion relation which can be derived from \eqref{appx MHD: eqn22} agree only up to order \(\mathcal{O}(k_y \rho_i)\) and disagree at higher orders. This is the reason why we had earlier calculated our time-dependent MHD model \eqref{MHD:eqn10} only up to this order.

\par
We have confirmed that in the case of zero shear, gyrokinetics and MHD agree at least to order \(\mathcal{O}(k_y \rho_i)\). This gives us a basis to presume that when the shear effects are weak, we expect there to be relatively good agreement between the two models. What remains is to derive a time-dependent gyrokinetic model with finite shear.

\subsection{Finite flow shear} \label{GK SECT: Finite shear} 

We now consider the full linear collisionless gyrokinetic equation \eqref{GK:eqn1} with the non-zero linear shear profile \(\boldsymbol{u} = S x \boldhat{y}\). As we have done in MHD, we perform the shearing box transformation \((t, \mathbf{r}) \rightarrow (t', \mathbf{r}')\) discussed in \S\ref{MHD SECT: shear box} and similarly for \((t,\boldsymbol{R}_s) \to (t', \boldsymbol{R}_s')\). The result is
\begin{equation}
    \begin{aligned}
		&
        \frac{\partial h_s}{\partial t'}  
		+
        \left(
        w_\parallel \boldhat{z} 
        +
        \frac{g \boldhat{y}}{\Omega_s} 
        \right)
        \cdot \frac{\partial h_s}{\partial \boldsymbol{R}_s'}
		=
		\frac{Z_s e F_{0s}}{T_s}
        \frac{\partial \left\langle  \delta\varphi  \right\rangle_{\boldsymbol{R}}}{\partial t'} 
		+
        \frac{c F_{0s}}{B}
        \left(
        \frac{\rmd \log n_s}{\rmd x} + \frac{m_s g}{T_s}
        \right) 
        \boldhat{y}\cdot
        \frac{\partial \langle \delta\varphi 
        \rangle_{\boldsymbol{R}}}{\partial \boldsymbol{R}_s'}
        .
	\end{aligned}
    \label{GK:eqn19}
\end{equation}
We note here that since this transformation is based on a local approximation, the gradient \(\rmd \log n_s/ \rmd x\) is a free parameter.

\par 
As was done in \eqref{MHD:eqn8} for MHD, we consider the following solutions for the perturbed quantities
\begin{align}
    &
    \delta\varphi (\boldsymbol{r}',t')
        = 
        \widehat{\delta\varphi}(t') \, e^{\im \boldsymbol{k}_\perp' \cdot \boldsymbol{r}'}
    ,
    \label{GK:eqn20}
    \\
    &
    h_s (\boldsymbol{R}_s', t')
        = 
        \hat{h}_s(t') e^{\im \boldsymbol{k}_\perp' \cdot \boldsymbol{R}_s'}
    ,
    \label{GK:eqn21}
\end{align}
where we have taken \(k_\parallel \rightarrow 0\) for flute modes. The gyroaverages are similar to \eqref{GK:eqn7} and \eqref{GK:eqn8}:
\begin{align}
    &
    \left\langle \delta\varphi \right\rangle_{\boldsymbol{R}}
        = 
        \widehat{\delta\varphi}(t') \, e^{\im \boldsymbol{k}_\perp'\cdot \boldsymbol{R}_s'}
        J_0 \left(a_s(t'')\right)
    ,
    \label{GK:eqn22}
    \\
    &
    \left\langle h_s \right\rangle_{\boldsymbol{r}}
        = 
        \hat{h}_s (t') e^{\im \boldsymbol{k}_\perp'\cdot \boldsymbol{r}'} J_0 \left(a_s(t'')\right)
    .
    \label{GK:eqn23}
\end{align}
Here, we have defined:
\begin{equation}
    \begin{aligned}
        a_s(t'') = \frac{w_\perp k_\perp (t')}{\Omega_s}
        = \frac{w_\perp}{\Omega_s} 
        \sqrt{k_x^2 + k_y^2} 
        =
        \frac{w_\perp}{\Omega_s} 
        \sqrt{(k_x' - S t' k_y')^2 + {k_y'}^2} 
        =
        \frac{w_\perp k_y'}{\Omega_s} \sqrt{1 + S^2 {t''}^2} 
        ,
    \end{aligned}
    \label{GK:eqn24}
\end{equation}
where the origin of time has been shifted i.e. \(t'' = t' - S^{-1} k_x'/k_y'\) such that for any value of \(k_x'\), \(t''=0\) always indicates the point where \(k_x = 0\) in the lab frame. Nevertheless, as we had done in MHD, we will consider the most unstable modes with \(k_x'=0\) which means this shifting is irrelevant, but we still include it here to be consistent with the theory developed in \citet{Alex12}.

\par 
We now rewrite the gyrokinetic system of equations \eqref{GK:eqn19} and \eqref{GK:eqn3} in terms of the new variables \((t'', \boldsymbol{k}')\), this gives us
\begin{align}
    &
        \left(
		\frac{\partial }{\partial t}  
		+
		\im \omega_{gs}
		\right)
		\hat{h}_{s} (t) 
		=
		\frac{Z_s e F_{0s}}{T_s}
		\left[
		\frac{\partial}{\partial t} 
		+
		\im (\omega_{gs}
		+
		\omega_{*s} )
		\right]
		J_0 \left(a_s(t) \right) \widehat{\delta\varphi} (t) 
        ,
    \label{GK:eqn25}
    \\
    &
        \sum_s \frac{Z_s^2 e^2 n_s}{T_s} \widehat{\delta\varphi} (t) 
		=
		\sum_s Z_s e \int \rmd^3\boldsymbol{w} \, 
		J_0 \left( a_s(t)\right)
        \hat{h}_{s}(t)  
        ,
    \label{GK:eqn26}
\end{align}
where the primes on all variables have been suppressed for clarity.

\subsubsection{Integral equation} \label{GK SECT: Finite shear - int eqn}

For finite shear, our dispersion relation takes the form of an integral equation:
\begin{equation}
    \begin{aligned}
        &\sum_s \frac{Z_s^2 e^2 n_s}{T_s}
        \left[ 
            1 - \Gamma_0(\lambda_s, \lambda_s)
        \right]
        \widehat{\delta\varphi}(t)
        =
        \sum_s Z_s e \int \rmd^3\boldsymbol{w} 
        J_0 \left(a_s(t) \right) e^{-\im \omega_{gs} t}
        \bigg[
            \hat{h}_s(0) 
        \\
        &- 
            \frac{Z_s e F_{0s}}{T_s} J_0(a_s(0)) \widehat{\delta\varphi}(0)
        \bigg]
       +
        \sum_s 
        \frac{Z_s^2 e^2 n_s}{T_s} \im \omega_{*s}
        \int_0^t \rmd\Delta t \, e^{- \im \omega_{gs} \Delta t} \Gamma_0(\lambda_s, \lambda_s')  \widehat{\delta\varphi}(t - \Delta t)
        \, .
    \end{aligned}
    \label{GK:eqn27}
\end{equation}
The derivation of this integral equation can be found in \autoref{appx GK int}. Here, we have defined \(\Gamma_0\) and \(a_s\) as
\begin{align}
    &
     \Gamma_0(\lambda_s, \lambda_s') 
     =
     \exp\left(- \frac{\lambda_s + \lambda_s'}{2} \right) 
     I_0 \left(\sqrt{\lambda_s \lambda_s'} \right)
     ,
    \label{GK:eqn28}
    \\
    &
    a_s(t) = \frac{w_\perp k_y}{\Omega_s} \sqrt{1 + S^2 t^2}
        = \frac{w_\perp \sqrt{2}}{v_{\text{th},s}}
    \sqrt{\frac{k_y^2 \rho_s^2 + S^2 t^2 k_y^2 \rho_s^2}{2}}
    =
    \frac{w_\perp}{\vths{s}} \sqrt{2\lambda_s(t)}
    .
    \label{GK:eqn29}
\end{align}
We introduce the parameter \(\lambda_s = k_\perp^2(t) \rho_i^2/2\) here to capture the evolution of the time-dependent wavenumber in the lab frame.

\par 
The first term on the RHS of equation \eqref{GK:eqn27} is just a source (\(t = 0\)) term necessary for initial value problems. It is the second term that captures the dynamics of the problem and is our focus analytically. However, when we solve \eqref{GK:eqn29} numerically later in \S\ref{SECT: numerical}, this source term will become pivotal in preventing us from obtaining a trivial solution.

\subsubsection{Short-time limit} \label{GK SECT: Finite shear - short time} 

Assuming that the source terms i.e. the \(t = 0\) term on the RHS of \eqref{GK:eqn27} can be neglected, we now consider the short-time regime \(St \ll 1\) earlier discussed in \S\ref{MHD SECT: short time}. In this regime, we have \(\lambda_s \approx \lambda_s' \approx \lambda_s(0) = k_y^2 \rho_s^2/2 \equiv \lambda_{s0}\). As a result, we have
\begin{equation}
    \begin{aligned}
        \Gamma_0(\lambda_s, \lambda_s)
        \approx 
        \Gamma_0(\lambda_s, \lambda_s')
        \approx
        \Gamma_0(\lambda_{s0}, \lambda_{s0} )
        =
        I_0(\lambda_{s0}) e^{-\lambda_{s0}}
        .
    \end{aligned}
    \label{GK:eqn30}
\end{equation}
The integral equation \eqref{GK:eqn27} then becomes
\begin{equation}
    \begin{aligned}
        &\sum_s \frac{Z_s^2 e^2 n_s}{T_s}
        \left[ 
            1 - \Gamma_0(\lambda_{s0}, \lambda_{s0})
        \right]
        \widehat{\delta\varphi}(t)
        \\
       &=
        \sum_s 
        \frac{Z_s^2 e^2 n_s}{T_s} \im \omega_{*s} \Gamma_0(\lambda_{s0}, \lambda_{s0})
        \int_0^t \rmd \Delta t \, e^{- \im \omega_{gs} \Delta t}   \widehat{\delta\varphi}(t - \Delta t)
        \, .
    \end{aligned}
    \label{GK:eqn30.1}
\end{equation}
Since we expect the effects of shear to be weak, we can assume time-independent growth rates i.e. \(\widehat{\delta\varphi}(t) \propto e^{-\im \omega t}\). We then obtain
\begin{equation}
    \begin{aligned}
        &\sum_s \frac{Z_s^2 e^2 n_s}{T_s}
        \left[ 
            1 - \Gamma_0(\lambda_{s0}, \lambda_{s0})
        \right]
        =
        \sum_s 
        \frac{Z_s^2 e^2 n_s}{T_s} 
		\omega_{*s} \Gamma_0(\lambda_{s0}, \lambda_{s0}) 
		\frac{e^{\im (\omega - \omega_{gs})t} - 1}{\omega - \omega_{gs} }
        \, .
    \end{aligned}
    \label{GK:eqn30.2}
\end{equation}
We would like to remove the oscillating term in \eqref{GK:eqn30.2} by somehow being able to have \(t\to\infty\) while respecting \(St\ll1\). A possible way to achieve this is to first, split the complex frequency into its real and imaginary parts i.e. \(\omega = \omega_R + \im \gamma\). We then have \(\lvert e^{\im(\omega - \omega_{gs})t}\rvert = e^{-\gamma t}\). We shall assume a small positive growth rate \(\gamma > 0\), which will be justified in a moment. Then by considering the limit where \(S\rightarrow 0\) and \(t\rightarrow \infty\), this would allow us to remove the oscillating term. 

\par 
For a hydrogen plasma, after using \eqref{GK:eqn30}, \eqref{GK:eqn30.2} then becomes
\begin{equation}
    \begin{aligned}
        \frac{1 - I_0(\lambda_i)e^{-\lambda_i}}{T_i}
        +
        \frac{1 - I_0(\lambda_e)e^{-\lambda_e}}{T_e}
        =
        - \frac{\omega_{*i}}{T_i} \frac{I_0(\lambda_i) e^{-\lambda_i}}{\omega - \omega_{gi}}
         - 
         \frac{\omega_{*e}}{T_e} \frac{I_0(\lambda_e) e^{-\lambda_e}}{\omega - \omega_{ge}}
         .
    \end{aligned}
    \label{GK:eqn31}
\end{equation}
Taking the long wavelength limit \(k_y\rho_i \ll 1\), the small mass ratio limit \(m_e/m_i \ll 1\), and assuming that \(T_i = T_e\), equation \eqref{GK:eqn31} reduces to the dispersion relation \eqref{GK:eqn13} of the zero shear case. 

\par 
This result justifies the earlier assumption that \(\gamma > 0\), since from \eqref{GK:eqn18}, we know to leading order that the growth rate is simply that of the ideal MHD interchange which is positive.

\subsubsection{Long time limit}

In \citet{Alex12}, a long-time limit (\(St \gg 1\)) asymptotic analysis for a similar integral equation was provided. Nevertheless, this analysis is difficult to replicate for the present problem due to the additional complexity of considering kinetic electrons. 

\par 
To understand this, we note that based on the definition of \(\lambda_s\) in \eqref{GK:eqn29}, we have \(\lambda_e = (m_e/m_i)\lambda_i\). The short time regime is unambiguous since we will consider modes in which \(k_y \rho_i \ll 1\) in which we have already seen that the dispersion relations from MHD and gyrokinetics that they agree in the limit of zero shear. Hence, the arguments of the Bessel functions \(\Gamma_0(\lambda_s,\lambda_s)\) are small regardless of whether the species index \(s\) refers to the ions or electrons.

\par 
The same cannot be said for the long time limit where \(St \gg 1\) and hence \(\lambda_s \approx S^2 t^2 k_y^2 \rho_i^2/2\). In \citet{Alex12}, this allowed for a large argument expansion of \(\Gamma_0(\lambda_s,\lambda_s)\). The issue here is that even though \(\lambda_i \gg 1\), it is not necessary that \(\lambda_e \gg 1\) due to the presence of the factor of the mass ratio for the electrons, thereby causing the long time regime to be ambiguous. 


\par 
The only definitive `long time regime' is one where \(St\) is so large that \(\lambda_e \gg 1\), as this automatically implies that \(\lambda_i \gg 1\). Therefore, there is no ambiguity when performing the large argument expansion of the Bessel functions for either species. In this scenario, we need to impose that
\begin{equation}
    \begin{aligned}
        \lambda_e = \frac{m_e}{m_i} \lambda_i 
        \gg 1
        .
    \end{aligned}
    \label{GK:eqn31.1}
\end{equation}
Since \(\lambda_i\) is a measure of the time-dependent \(k_\perp^2(t) \rho_i^2/2\) in the lab frame, the condition \eqref{GK:eqn31.1} then demands that:
\begin{equation}
    \begin{aligned}
        k_\perp(t) \rho_i \gg \sqrt{\frac{2m_i}{m_e}} \approx 60
        ,
    \end{aligned}
    \label{GK:eqn31.2}
\end{equation}
which is well outside the regime of interest. 

\par 
The reader may also wonder if an analysis in the intermediate regime, where \(\lambda_i \gg 1\) but \(\lambda_e \sim 1\), is possible. This would require us to borrow ideas from \S3.2 in \citet{Alex12} and seek solutions of the form:
\begin{equation}
    \begin{aligned}
        \widehat{\delta\varphi} (t)
        =
        \widehat{\delta\varphi} \exp \left[
            \int_0^t \rmd t' \, \gamma(t')
        \right]
        .
    \end{aligned}
    \label{GK:eqn31.3}
\end{equation}
For \(\widehat{\delta\varphi}(t-\Delta t)\) in our integral equation, we need to set up an expansion by considering \(\Delta t \ll t\). This is equivalent to requiring a limited time history, where for any time \(t\), the contributions to the integral are largely weighted by its immediate temporal neighbourhood. For the ionic integral, because \(\lambda_i \gg 1\), this is achieved by a large argument expansion of \(\Gamma_0(\lambda_i, \lambda_i')\). However, the same expansion is not possible for the electron integral since \(\lambda_e \sim 1\). 

\par 
Even if the expansion for the electron integral is possible, we would need to deal with the issue of truncating the expansion of \(\widehat{\delta\varphi}(t-\Delta t)\). While this is possible in \citet{Alex12} by assuming all derivatives of \(\gamma(t)\) to be small, we do not believe this is an assumption we can make based on our later numerical results in  \S\ref{SECT: numerical}.

\par 
In conclusion, outside the short time regime, we believe it is best to numerically integrate \eqref{GK:eqn27}.

\subsection{Volterra integral equation from gyrokinetics} \label{GK SECT: Volterra}

We would now like to numerically integrate \eqref{GK:eqn27}; however, as previously mentioned, we cannot neglect the source term on the RHS of the equation, as this will inevitably result in the trivial solution. As this is an initial value problem, the choice of the source term will affect the final solution. As we eventually want to compare with GX, which automatically chooses the initial density perturbation to be a perturbed Maxwellian, we shall choose the source term to be:
\begin{equation}
    \begin{aligned}[b]
        \hat{h}_s(0) - \frac{Z_s e F_{0s}}{T_s} J_0 (a_s(0)) \widehat{\delta\varphi}(0)
        =
        \kappa F_{0s}
        =
        \kappa \frac{n_s }{\pi^{3/2}\vths{s}^3}e^{-w^2/\vths{s}^2}
        ,
    \end{aligned}
    \label{GK:eqn32}
\end{equation}
%
for some sufficiently small value of \(\kappa\). This choice ensures both GX and the gyrokinetic integral equation \eqref{GK:eqn27} produce the same exponential solution when \(S = 0\). 

\par 
Substituting \eqref{GK:eqn32} into equation \eqref{GK:eqn27}, we obtain:
\begin{equation}
    \begin{aligned}
        &\sum_s \frac{Z_s^2 e^2 n_s}{T_s}
        \left[ 
            1 - \Gamma_0(\lambda_s, \lambda_s)
        \right]
        \widehat{\delta\varphi}(t)
        =
        \kappa \sum_s  Z_s e n_s 
        e^{-\im\omega_{gs}t} e^{-\lambda_s(t)/2}
        \\
        &+\sum_s 
        \frac{Z_s^2 e^2 n_s}{T_s} \im \omega_{*s}
        \int_0^t \rmd \Delta t \, e^{- \im \omega_{gs} \Delta t} \Gamma_0(\lambda_s, \lambda_s')  \widehat{\delta\varphi}(t - \Delta t)
        ,
    \end{aligned}
    \label{GK:eqn33}
\end{equation}
where we have utilised the integral identity \citep{AbramowitzStegun}:
\begin{equation}
    \begin{aligned}
        \int_0^\infty \rmd t \, e^{-a^2 t^2} t^{\nu + 1} J_\nu (bt)  
        =
        \frac{b^\nu}{(2a^2)^{\nu +1}} e^{-b^2/4a^2}
        .
    \end{aligned}
    \label{GK:eqn34}
\end{equation}
For the case of a hydrogen plasma, performing the species expansion, undoing the coordinate transformation \(t' \rightarrow \Delta t\) described in \eqref{appx GK int: eqn4}, using the relations \eqref{GK:eqn14}--\eqref{GK:eqn16} to write everything in ionic terms, and assuming \(T_i = T_e\), we finally get
\begin{equation}
    \begin{aligned}
        &
        \left[ 
            2 - \Gamma_0(\lambda_i, \lambda_i)
             - \Gamma_0\left(\lambda_i \frac{m_e}{m_i}, \lambda_i \frac{m_e}{m_i}\right)
        \right]
        \widehat{\delta\varphi}(t)	
	    =
        \frac{\kappa T_i}{e} 
        \left(
        e^{-\lambda_i / 2} 
        e^{-\im\omega_{gi}t}
    	-
    	    e^{-\lambda_i / 2 \frac{m_e}
        {m_i}}
        e^{\im\frac{m_e}{m_i}\omega_{gi} t}
        \right)
        \\
        & 
        +
         \im \omega_{*i}
         \int_0^t \rmd t'\,
         \left[ 
             e^{- \im \omega_{gi} (t-t')} \Gamma_0(\lambda_i, \lambda_i') 
    	      -
             e^{\im \frac{m_e}{m_i}\omega_{gi} (t-t')} \Gamma_0\left(\lambda_i \frac{m_e}{m_i}, \lambda_i' \frac{m_e}{m_i} \right)  
        \right]\widehat{\delta\varphi}(t')
        ,
    \end{aligned}
    \label{GK:eqn35}
\end{equation}
Self-consistently, at \(t=0\), we must have
\begin{equation}
    \begin{aligned}
        &
        \widehat{\delta\varphi}(0) = 
        \frac{\kappa T_i}{e} 
        \frac{
        	e^{-\lambda_{i0}/2} - e^{- \frac{m_e}{m_i} \lambda_{i0}/2}
        }{
        	2 - \Gamma_0(\lambda_{i0}, \lambda_{i0})
                     - \Gamma_0\left(\lambda_{i0} \frac{m_e}{m_i}, \lambda_{i0} \frac{m_e}{m_i}\right)
        }
        .
    \end{aligned}
    \label{GK:eqn36}
\end{equation}
We now divide \eqref{GK:eqn35} throughout by \(\widehat{\delta\varphi}(0)\), which gives us:
\begin{equation}
    \begin{aligned}
        &\frac{\widehat{\delta\varphi}(t)}{\widehat{\delta\varphi}(0)}
        =
    	\frac{
    		e^{-\lambda_i / 2} e^{-\im\omega_{gi} t}
    		-
    		e^{-\lambda_i / 2 \frac{m_e}{m_i}}
    		e^{\im \frac{m_e}{m_i}\omega_{gi} t}
    	}{
    		e^{-\lambda_{i0}/2} - e^{-\frac{m_e}{m_i} \lambda_{i0}/2}
    	}
    	\frac{
    		2 - \Gamma_0(\lambda_{i0}, \lambda_{i0})
    	             - \Gamma_0\left(\lambda_{i0} \frac{m_e}{m_i}, \lambda_{i0} \frac{m_e}{m_i}\right)
    	}{
    		2 - \Gamma_0(\lambda_i, \lambda_i)
                 - \Gamma_0\left(\lambda_i \frac{m_e}{m_i}, \lambda_i \frac{m_e}{m_i}\right)
    	}
        \\
        & +
         \im \omega_{*i}
        \int_0^t \rmd t'\, 
		\frac{
			e^{- \im \omega_{gi} (t-t')} \Gamma_0(\lambda_i, \lambda_i') 
			-
				e^{\im \frac{m_e}{m_i} \omega_{gi} (t-t')} \Gamma_0\left(\lambda_i  \frac{m_e}{m_i}, \lambda_i' \frac{m_e}{m_i} \right)  
		}{
			2 - \Gamma_0(\lambda_i, \lambda_i)
             - \Gamma_0\left(\lambda_i \frac{m_e}{m_i}, \lambda_i \frac{m_e}{m_i}\right)
			 }
	\frac{\widehat{\delta\varphi}(t')}{\widehat{\delta\varphi}(0)}
    .
    \end{aligned}
    \label{GK:eqn37}
\end{equation}
This is a \textit{Volterra equation of the second kind} \citep{LinzVolterra} in which we can solve numerically.\footnote{The numerical solver used can be found here: \url{https://github.com/ianabel/volterra}.} It is the solution from this equation which we hope to replicate with GX.

\section{Numerical results}\label{SECT: numerical}

So far, we have derived both an MHD and an analytical gyrokinetic model for the gravitational flute interchange in \S\ref{SECT: MHD gyro} and \S\ref{SECT: GK}, respectively. We would like to compare the predictions between both models, as well as the GX gyrokinetic code \citep{Noah18, Noah24}; the latter would allow for the theory discussed in this paper to be used as a linear flow shear benchmark for GX. We note that we are using an alternate branch of GX that includes flow shear as well as gravity.\footnote{This branch of GX can be found at \url{https://bitbucket.org/gyrokinetics/gx/branch/gravity} and is based on \url{https://bitbucket.org/gyrokinetics/gx/branch/next}.}

\subsection{GX Normalisations}

\begin{table}
    \centering
    \setlength{\tabcolsep}{8pt}
    \begin{tabular}{l c c c c c}
        \textbf{Quantity:}
        &
        \(\omega, \omega_{gs}, \omega_{*s}, S, t\)
        &
        \(k_y\)
        &
        \(m_s\)
        &
        \(g\)
        &
        \(\rmd \log n_s/ \rmd x\)
        \\[0.5em]
        \textbf{Normalised by:}
        &
        \(\vths{\text{ref}}/a\)
        &
        \(\rho_{\text{ref}}\)
        &
        \(m_\text{ref}\)
        &
        \(a/\vths{\text{ref}}^2\)
        &
        \(a\)
    \end{tabular}
    \captionsetup{width = 0.9\textwidth}
    \caption{Normalisation constants following the GX convention. Note that GX requires a definition of a reference species denoted by the `ref' subscript. In our case, the reference species is simply the ions.}
    \label{table1}
\end{table}

To make direct comparisons with GX, it is convenient for us to normalise both the MHD and gyrokinetic equations for the case of zero shear, \eqref{MHD:eqn5} and \eqref{GK:eqn12}, and the case of finite shear, i.e. \eqref{MHD:eqn10} and \eqref{GK:eqn37}. We shall follow the normalisations found in Appendix A of \citet{Noah24} and, in particular, those depicted in \autoref{table1}. It is unimportant to know the normalisations for \(\delta n_s\) and \(\delta\varphi\) since we are only interested in their ratios to their initial values. We note here that this reference follows a different convention in the definition of \(\vths{s}\) (and hence \(\rho_s\)), in which the factor of \(\sqrt{2}\) is absent. Nevertheless, our derived equations can be easily modified to include this factor. From this point on, for any quantity \(\alpha\), we shall denote its corresponding GX normalised quantity with a tilde above, i.e. \(\tilde{\alpha}\). 

\par 
In the case of zero shear, the normalised MHD \eqref{MHD:eqn5} and gyrokinetic \eqref{GK:eqn12} dispersion relations are:
\begin{align}
    &
    \tilde{\omega} 
    -
    \frac{1}{2} \left[
        \left(
            \tilde{\omega}_{gi} + \tilde{\omega}_{*i} 
        \right)
        \pm 
        \sqrt{
            \left(\tilde{\omega}_{gi} + \tilde{\omega}_{*i}\right)^2
            -
            4 \frac{\tilde{\omega}_{gi} \tilde{\omega}_{*i}}{\tilde{k}_y^2}
        }
    \right]
    =
    0
    ,
    \label{numerical:eqn1}
    \\
    &
    2-
	\frac{\tilde{\omega} - \tilde{\omega}_{gi} - \tilde{\omega}_{*i}}{\tilde{\omega} - \tilde{\omega}_{gi}} 
	I_0 \left(\tilde{k}_y^2 \right)
	e^{-\tilde{k}_y^2 }
	+
	\frac{\tilde{\omega} + \tilde{m}_e\tilde{\omega}_{gi} + \tilde{\omega}_{*i}}{\tilde{\omega} + \tilde{m}_e\tilde{\omega}_{gi}} 
	I_0 \left(\tilde{m}_e\tilde{k}_y^2  \right)
	e^{-\tilde{m}_e\tilde{k}_y^2 }
    =
    0
    .
    \label{numerical:eqn2}
\end{align}
In the case of finite shear, the time-dependent MHD \eqref{MHD:eqn10} and gyrokinetic \eqref{GK:eqn37} equations are:
\begin{align}
    &
    \begin{aligned}
        &
        \left\{
        \frac{\partial^2 }{\partial \tilde{t}^2} 
    	+
    	\Bigg[
    		\frac{2 \tilde{S}^2 \tilde{t}}{\tilde{S}^2 \tilde{t}^2 + 1}
    		- 
    		\im( \tilde{\omega}_{gi} + \tilde{\omega}_{*i})
    	\Bigg]
    	\frac{\partial}{\partial \tilde{t}} 
    	-
        \frac{
            \im\tilde{S}^2 \tilde{t}\left(2\tilde{\omega}_{gi} + \tilde{\omega}_{*i}\right) 
            +
            \tilde{\omega}_{gi}\tilde{\omega}_{*i}/\tilde{k}_y^2
        }{\tilde{S}^2 \tilde{t}^2 + 1}
        \right\}
    	\frac{\widetilde{\delta n}_i(\tilde{t})}{\widetilde{\delta n}_i(0)}
    	=
    	0
        ,
    \end{aligned}
    \label{numerical:eqn3}
    \\
    \intertext{and}
    &
    \begin{aligned}
        &\frac{\widetilde{\delta\varphi}(\tilde{t})}{\widetilde{\delta\varphi}(0)}
        =
    	\frac{
    		e^{-\lambda_i / 2} e^{-\im\tilde{\omega}_{gi} \tilde{t}}
    		-
    		e^{-\tilde{m}_e \lambda_i / 2 }
    		e^{\im \tilde{m}_e \tilde{\omega}_{gi} \tilde{t}}
    	}{
    		e^{-\lambda_{i0}/2} - e^{-\tilde{m}_e \lambda_{i0}/2}
    	}
    	\frac{
    		2 - \Gamma_0(\lambda_{i0}, \lambda_{i0})
    	             - \Gamma_0\left(\tilde{m}_e \lambda_{i0} , \tilde{m}_e \lambda_{i0} \right)
    	}{
    		2 - \Gamma_0(\lambda_i, \lambda_i)
                 - \Gamma_0\left(\tilde{m}_e \lambda_i , \tilde{m}_e\lambda_i \right)
    	}
        \\
        & \hspace{1.5cm} +
         \im \tilde{\omega}_{*i}
        \int_0^{\tilde{t}} \rmd \tilde{t}'\, 
		\frac{
			e^{- \im \tilde{\omega}_{gi} (\tilde{t}-\tilde{t}')} \Gamma_0(\lambda_i, \lambda_i') 
			-
				e^{\im \tilde{m}_e \tilde{\omega}_{gi} (\tilde{t}-\tilde{t}')} \Gamma_0\left(\tilde{m}_e \lambda_i, \tilde{m}_e\lambda_i' \right)  
		}{
			2 - \Gamma_0(\lambda_i, \lambda_i)
             - \Gamma_0\left(\tilde{m}_e\lambda_i , \tilde{m}_e\lambda_i \right)
			 }
	\frac{\widetilde{\delta\varphi}(\tilde{t}')}{\widetilde{\delta\varphi}(0)}
    .
    \end{aligned}
    \label{numerical:eqn4}
\end{align}
Note that, in \eqref{numerical:eqn4}, using the definition \eqref{GK:eqn29}, we have \(\lambda_i(\tilde{t}) = \tilde{k}_y^2  (1 + \tilde{S}^2 \tilde{t}^2)\). 

\par 
In the equations \eqref{numerical:eqn1}--\eqref{numerical:eqn4}, the free parameters are given by \(\{\tilde{\omega}_{gi}, \tilde{\omega}_{*i}, \tilde{k}_y, \tilde{S}\}\). However, the input parameters in GX are \(\{\tilde{g}, f_\text{prim}, \tilde{k}_y, \tilde{S}\}\), where \(\tilde{g}\) and \(f_\text{prim}\) are the normalised gravity and density gradient respectively. We have elected to explicitly denote the latter as \(f_{\text{prim}}\) because this is the actual input parameter in GX. These two sets of free parameters are then related by
\begin{align}
    &
    \tilde{\omega}_{gi} = \tilde{g}\tilde{k}_y
    ,
    \label{numerical:eqn5}
    \\
    \intertext{and}
    &
    \tilde{\omega}_{*i} = \tilde{k}_y f_\text{prim}
    .
    \label{numerical:eqn6}
\end{align}

\subsection{Gyrokinetic perturbed density}

For finite shear, we note that the gyrokinetic integral equation \eqref{numerical:eqn4} solves for \(\widetilde{\delta\varphi}/\widetilde{\delta\varphi}(0)\), while the MHD equation \eqref{numerical:eqn3} solves for \(\widetilde{\delta n}_i/\widetilde{\delta n}_i(0)\). We therefore need to find a way to solve for \(\widetilde{\delta n}_i/\widetilde{\delta n}_i(0)\) in the former. In \autoref{appx GK density}, we demonstrate how this can be achieved by using our earlier solution for \(\hat{h}_i(t)\) to compute \(\widehat{\delta n}_i(t)\) from \(\widehat{\delta\varphi}(t)\). After which, applying the GX normalisations would give us \(\widetilde{\delta n}_i\). The result is:
\begin{equation}
    \begin{aligned}
	&\frac{\widetilde{\delta n}_i }{\widetilde{\delta n}_i (0)}
    =
    \im \tilde{\omega}_{*i}  
    \int_0^{\tilde{t}} \rmd \tilde{t}' \, 
	\frac{
		e^{-\im \tilde{\omega}_{gi} (\tilde{t}-\tilde{t}')} 
	    \Gamma_0(\lambda_i, \lambda_i')
	}{
		\alpha_0
	}
	\frac{\widetilde{\delta\varphi} (\tilde{t}')}{\widetilde{\delta\varphi} (0)}
    \\
	&
	+
	\frac{1}{\alpha_0}
    \left\{
        \left[ \Gamma_0(\lambda_i, \lambda_i) - 1\right]
		\frac{\widetilde{\delta \varphi}}{\widetilde{\delta\varphi} (0)}
		+
		  e^{-\im \tilde{\omega}_{gi} \tilde{t}}  e^{-\lambda_i/2} 
		\frac{
			2 - \Gamma_0(\lambda_{i0}, \lambda_{i0})
					 - \Gamma_0\left(\tilde{m}_e \lambda_{i0} , \tilde{m}_e \lambda_{i0} \right)
		}{
			e^{-\lambda_{i0}/2} - e^{- \tilde{m}_e \lambda_{i0}/2}
		}
    \right\}
    ,
\end{aligned}
    \label{numerical:eqn7}
\end{equation}
where
\begin{equation}
    \begin{aligned}
	   \alpha_0
       &=
       e^{-\lambda_{i0}/2} 
		\frac{
			2 - \Gamma_0(\lambda_{i0}, \lambda_{i0})
					 - \Gamma_0\left(\tilde{m}_e\lambda_{i0}, \tilde{m}_e\lambda_{i0} \right)
		}{
			e^{-\lambda_{i0}/2} - e^{- \tilde{m}_e\lambda_{i0}/2}
		}
		 +
		 \left[
			 \Gamma_0(\lambda_{i0}, \lambda_{i0}) - 1 
		 \right]
        .
    \end{aligned}
    \label{numerical:eqn8}
\end{equation}
In other words, once \eqref{numerical:eqn4} is solved, the solution must be inserted into \eqref{numerical:eqn7}. The time-integral can then be evaluated by numerical integration where we have opted for the trapezoidal rule. This result is then the correct quantity to be compared with MHD. In \autoref{appx GK density} we will also explain how to obtain this quantity from GX.

\subsection{Influence of initial conditions}\label{SECT: numerical - init cond} 

In order to choose appropriate values for the free parameters \(\tilde{\omega}_{gi}\) and \(\tilde{\omega}_{*i}\), we will need to analyse the structure of the (finite shear) equations. In particular, the source term in the gyrokinetic integral equation \eqref{numerical:eqn4} is of considerable interest, which we recall is a consequence of the choice of an initial density perturbation \eqref{GK:eqn32}.

\par 
Firstly, we note that this source term is the only term that does not contain the perturbation ratio \(\widetilde{\delta\varphi}/\widetilde{\delta\varphi}(0)\) in the integral equation \eqref{numerical:eqn4}. On the other hand, no such term exists in the MHD equation \eqref{numerical:eqn3} as all terms are multiplied by \(\widetilde{\delta n_i}/\widetilde{\delta n_i}(0)\). While it is possible to capture the same initial conditions in the MHD model via the initial value of its derivative (see \autoref{appx MHD init cond}), we should not expect the time evolution of this source term to be fully captured. This boils down to the fact that, even though we have set \eqref{GK:eqn32} as a condition on \(\hat{h}_s\), from \eqref{GK:eqn6}, we see that \(h_s(0) \propto e^{\im k_y \boldsymbol{R}_s\cdot\boldhat{y}}\). Physically, this means that finite FLR effects are embedded into the initial conditions of the gyrokinetic solution and the MHD model cannot hope to fully capture it as it only contains FLR effects to first order. This leads to a natural divergence, which we will later demonstrate in our numerics when source term effects are strong. 

\par 
Secondly, we would like to determine when the source term is dominant. This is important because we recall from our prior derivations that both the gyrokinetic integral equation and MHD equation arrive at the same dispersion relation \eqref{GK:eqn18} in the short-time limit \(St \ll 1\) provided that the source-term is neglected in the former. This implies that the main FLR physics is in fact captured solely by the integral term of \eqref{numerical:eqn4}. Since it is the only term multiplied by \(\tilde{\omega}_{*i}\), we can make it dominant by choosing \(\tilde{\omega}_{*i}\) to be sufficiently large. In our numerics, we shall illustrate two case \(f_\text{prim} = 0.5\) and \(f_{\text{prim}} = 15\) where the former is termed the \textit{strong source regime} and the latter is termed the \textit{weak source regime}.

\par 
Thirdly, we note that the parameter \(\tilde{\omega}_{gi}\) is tied to oscillatory terms in \eqref{numerical:eqn4}. Therefore, to avoid significant oscillation, we choose a `safe' value of \(\tilde{g} = 0.05\). In addition, if we compare our MHD equation \eqref{MHD:eqn10} to that of \citet{Ng05} given in \eqref{appx Ng MHD:eqn3}, we observe that the difference between these two equations is solely due to the terms multiplied by a factor of \(\tilde{\omega}_{gi}\). Therefore, by choosing \(\tilde{g}\) and hence \(\tilde{\omega}_{gi}\) to be small, we can ensure that the solutions from both MHD models are numerically similar.

\subsection{Zero flow shear} \label{numerical SECT: zero shear}

\begin{figure}
  \begin{minipage}[t]{0.48\textwidth}
    \centering
    \(f_{\text{prim}} = 0.5\) 
    \\
    \includegraphics[scale = 0.5]{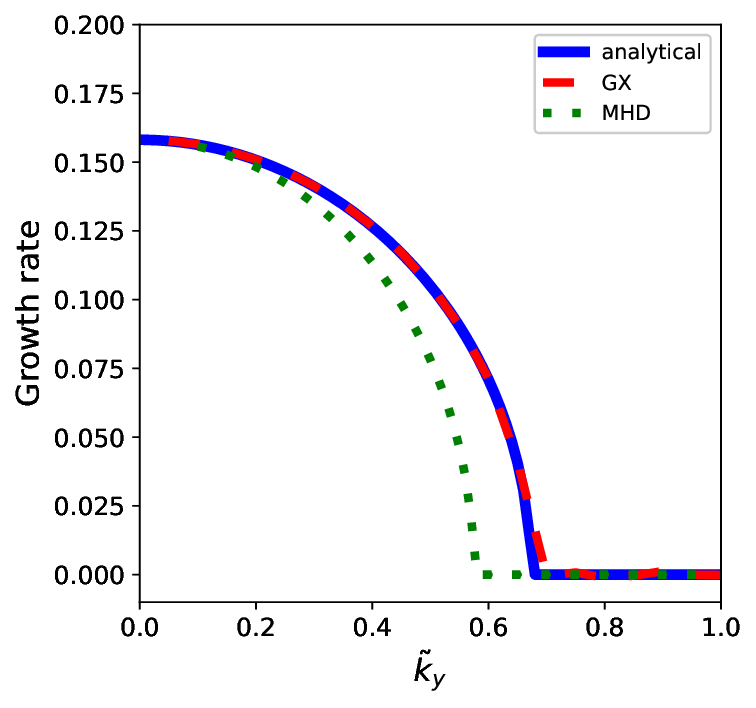} 
  \end{minipage}\hfill
  \begin{minipage}[t]{0.48\textwidth}
    \centering
    \(f_{\text{prim}} = 15\) 
    \\
    \includegraphics[scale = 0.5]{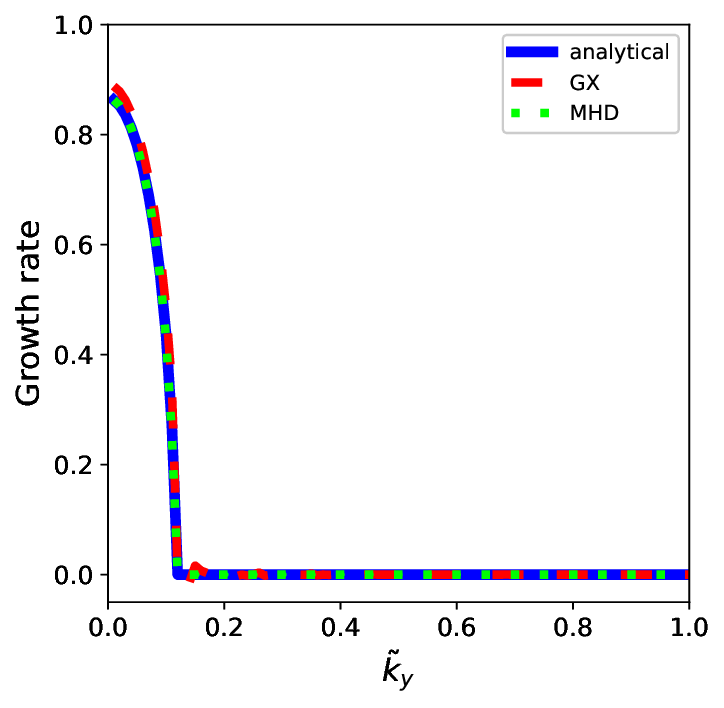}
  \end{minipage}
  \captionsetup{width = 0.9\textwidth}
    \caption{
     Plot of the growth rate \(\Im\{\omega\}\) against \(\tilde{k}_y\) computed from MHD \eqref{numerical:eqn1}, the analytical gyrokinetic solution \eqref{numerical:eqn2}, and the gyrokinetic code GX. The left panel represents the strong source regime while the right panel represents the weak source regime.
    }
    \label{numerical:fig1}
\end{figure}

We first compare the linear growth rates predicted by MHD, gyrokinetics, and GX in the case of zero shear. This is depicted in \autoref{numerical:fig1} and it is evident that the unstable modes are those for which \(\tilde{k}_y \ll 1\). In the case of finite shear, the shear pushes \(\tilde{k}_\perp\) (equivalently \(\tilde{k}_y\) for zero shear) higher and higher until these unstable modes reach the FLR stabilisation threshold, beyond which we expect a decaying solution. Therefore, by comparing the predictions of this threshold value by these different models, we would be able to get a sense of when the decay should be observed.

\par
In the strong source regime, we see that MHD predicts the FLR stabilisation threshold to be \(\tilde{k}_\perp = \tilde{k}_y \approx 0.575\), while gyrokinetics predicts it to be \(\tilde{k}_\perp = \tilde{k}_y \approx  0.674\). Later on, for finite shear, since \(\tilde{k}_\perp^2 = \lambda_i(\tilde{t}) = \tilde{k}_y^2 (1 + \tilde{S}^2 \tilde{t}^2)\), we expect a decay at:
\begin{align}
    &
    \tilde{t}_{\text{FLR}, \text{MHD}} 
    \approx 
    \sqrt{\frac{1}{\tilde{S}^2} \left(\frac{0.575^2}{\tilde{k}_y^2} - 1\right)}
    ,
    \label{numerical:eqn9}
    \\
    &
    \tilde{t}_{\text{FLR}, \text{GK}} 
    \approx 
    \sqrt{\frac{1}{\tilde{S}^2} \left(\frac{0.674^2}{\tilde{k}_y^2} - 1\right)}.
    \label{numerical:eqn10}
\end{align}

\par 
In the weak source regime, both MHD and gyrokinetics predict the same FLR stabilisation threshold to be \(\tilde{k}_\perp = \tilde{k}_y \approx 0.115\). Hence, for finite shear, the approximate time a decay solution is expected is:
\begin{equation}
    \begin{aligned}
        \tilde{t}_{\text{FLR}, \text{MHD}} 
        =
        \tilde{t}_{\text{FLR}, \text{GK}}
        \approx
        \sqrt{\frac{1}{\tilde{S}^2} \left(\frac{0.115^2}{\tilde{k}_y^2} - 1\right)}.
    \end{aligned}
    \label{numerical:eqn11}
\end{equation}

\par  
Notice that the threshold wavenumber is much smaller in the strong source regime compared to the weak source regime. However, we recall from the derivation of the zero shear gyrokinetic dispersion relation that there is no `source term'. It only presents itself in the finite shear case where we did not assume normal mode time-dependence. This implies that at zero shear, the system does not know whether it is being affected by initial conditions or not. It only understands that the instability is weak for \(f_\text{prim}=0.5\) and strong for \(f_{\text{prim}} = 15\). This gives us the interesting physical result that \textit{weaker instabilities take longer to FLR stabilise.} This will later be apparent in our numerical simulations.

\subsection{Finite flow shear}

We now turn our attention to the time traces of the ratio to the initial perturbation predicted by MHD \eqref{numerical:eqn3} and gyrokinetics \eqref{numerical:eqn7} in the presence of flow shear. Before discussing the numerical results, we would first like to describe a few parameters that one should pay attention to when running these simulations in GX.

\subsubsection{GX inputs for box size and number of modes in the x direction}

The maximum time that our flow shear simulation can run in GX is controlled by the following input parameters: the shear parameter \(\tilde{S}\), the number of Fourier modes in \(\tilde{k}_x\) which we shall denote by \(N_{k_x}\), and the box size in \(x\) which we shall denote by \(\tilde{x}_0\). To understand this dependence, we note that \(N_{k_x}\) sets both the number of positive and negative \(\tilde{k}_x\) modes, symmetric about zero. On the other hand, the \(\tilde{k}_x\) interval is determined by \(\Delta \tilde{k}_x = 1/\tilde{x}_0\). This sets the (absolute) maximum \(\tilde{k}_x\) in the system to be \(\tilde{k}_{x,\max} = \Delta \tilde{k}_x (N_{k_x}-1)/2\). As \(\tilde{k}_x\) is increasing with time in our problem as seen in \eqref{MHD:eqn9}, we can calculate when \(\lvert \tilde{k}_x(t)\rvert\) reaches \(\tilde{k}_{x,\max}\), beyond which it falls off the grid:
\begin{equation}
    \begin{aligned}
        \lvert \tilde{k}_{x,\max} (\tilde{t}_{\max}) \rvert 
        =
        \tilde{S} \tilde{t}_{\max}  \tilde{k}_y
        \Longrightarrow 
        \tilde{t}_{\max} 
        =
        \frac{\Delta \tilde{k}_x}{\tilde{S} \tilde{k}_y }
         \frac{N_{k_x}-1}{2}
         =
         \frac{N_{k_x}-1}{2\tilde{x}_0 \tilde{S} \tilde{k}_y }
         .
    \end{aligned}
    \label{numerical:eqn12}
\end{equation}
This is important for two reasons. Firstly, when analysing scenarios with strong flow shear, one needs either a large \(N_{k_x}\) or a small \(\tilde{x}_0\) to allow the simulation to run for a reasonable amount of time. Secondly, our numerical tests reveal that choosing too small a value of \(\tilde{x}_0\) would result in unphysical discontinuities in the solution.\footnote{This is a minor bug in GX and only manifests in simulations with kinetic electrons.} Therefore, some care must be taken when choosing the inputs \(N_{k_x}\) and \(\tilde{x}_0\). 

\par 
We will work with \(\tilde{x}_0 = 256.0\) in our flow shear simulations and use \(N_{k_x} = 2^{15}\). An exception to this is the weak shear scenario for the strong source regime, where a longer simulation time is required to demonstrate the decay, imploring us to use \(N_{k_x} = 2^{16}\).

\par 
GX is a parallelised code with a Fourier decomposition in (perpendicular) configuration space and Hermite-Laguerre decomposition for velocities \citep{Noah18, Noah24}. It works with a maximum of 1 GPU per species per Hermite moment; therefore, as these simulations used 8 Hermite moments and 2 kinetic species, 4 Perlmutter nodes with 4 GPUs per node were utilised.\footnote{Perlmutter is the high performance computing facility at the National Energy Research Scientific Computing Center.}

\subsubsection{Observations for increasing flow shear}

\begin{figure}
    \centering
    
    \begin{subfigure}{0.48\textwidth}
        \centering
        \large \(f_\text{prim}=0.5\) \normalsize 
        \\
        \includegraphics[width=\linewidth]{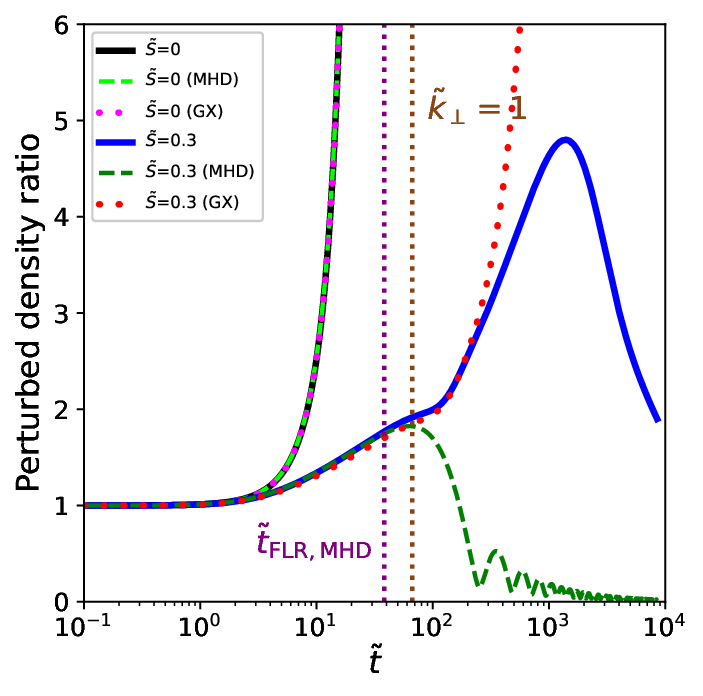}
    \end{subfigure}\hfill
    \begin{subfigure}{0.48\textwidth}
        \centering
        \large \(f_\text{prim}=15\) \normalsize
        \\
        \includegraphics[width=\linewidth]{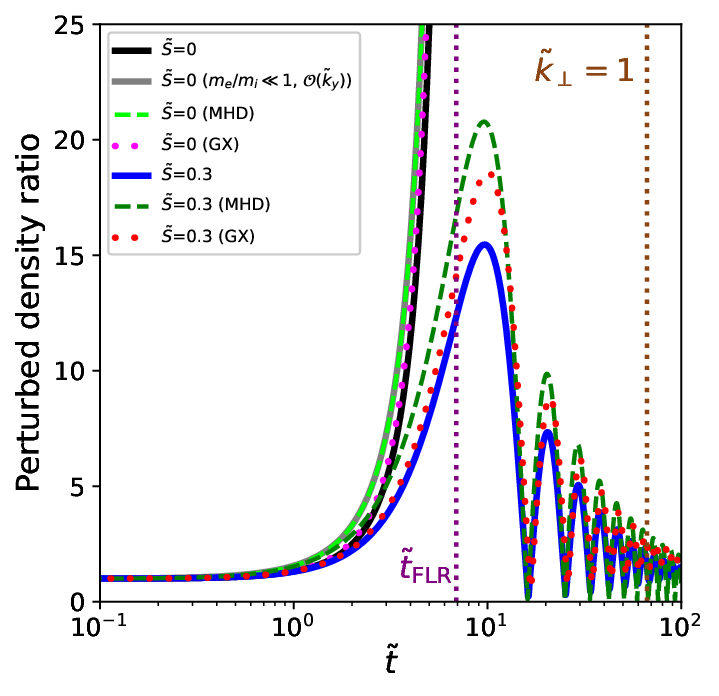}
    \end{subfigure}
    
    \smallskip
    
    \begin{subfigure}{0.48\textwidth}
        \centering
        \includegraphics[width=\linewidth]{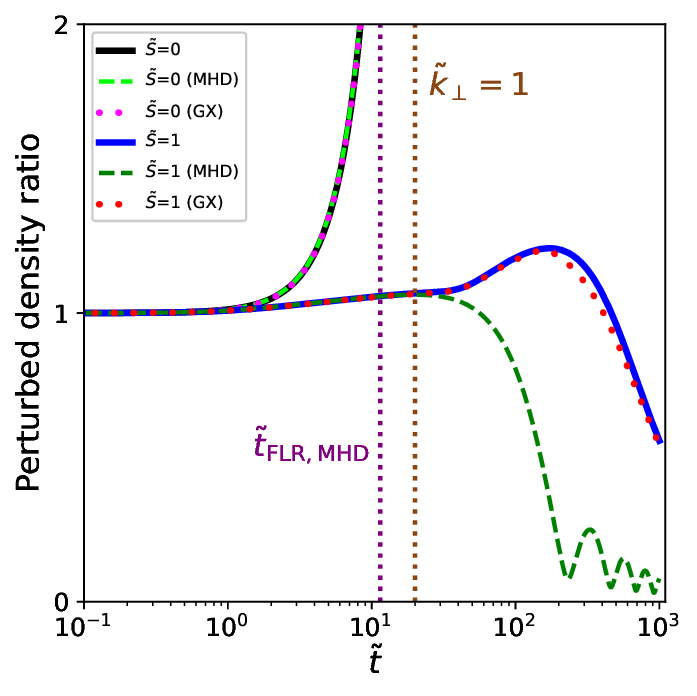}
    \end{subfigure}\hfill
    \begin{subfigure}{0.48\textwidth}
        \centering
        \includegraphics[width=\linewidth]{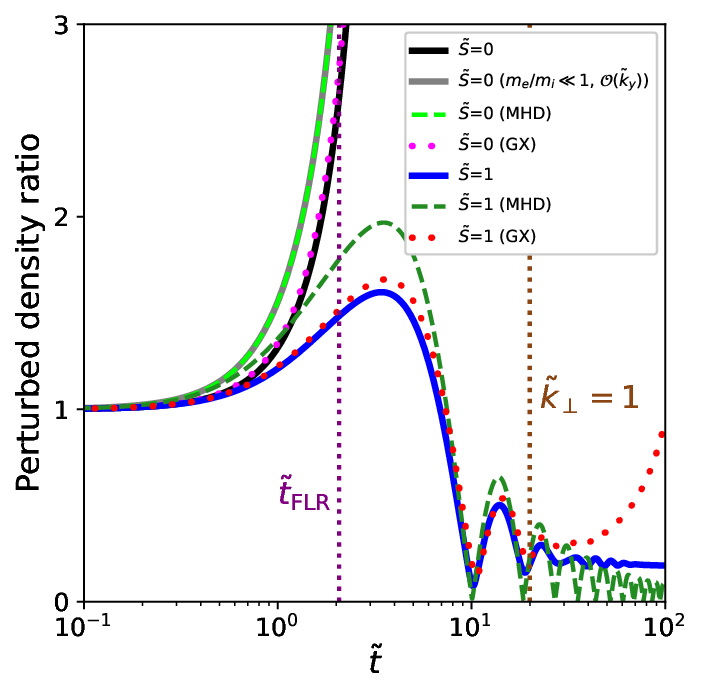}
    \end{subfigure}
    \captionsetup{width = 0.9\textwidth}
    \caption{
        Plots of \(\widetilde{\delta n}_i / \widetilde{\delta n}_i(0)\) against \(\tilde{t}\). The left column represents the strong source regime \(f_{\text{prim}} = 0.5\) in the case of weak shear (\(\tilde{S}=0.3\)) and strong shear (\(\tilde{S} = 1\)). This is the same for weak source regime \(f_\text{prim}=15\) on the right column. In each figure, the \(\tilde{S}=0\) solution for each model is always presented as a reference. The black and blue solid lines represent gyrokinetic solutions computed analytically from \eqref{numerical:eqn7} and \eqref{numerical:eqn4}. The grey solid line represent a \(m_e/m_i \ll 1\) and \(\mathcal{O}(\tilde{k}_y)\) expansion of the gyrokinetic solution as described in \autoref{appx GK low order}. The lime and green dashed lines are from the MHD equation \eqref{numerical:eqn3}. The pink and red dotted lines are the solutions from GX. The vertical dotted purple line represents the supposed time in which FLR stabilisation, as predicted by MHD, is expected to occur, calculated from \eqref{numerical:eqn9} and \eqref{numerical:eqn11} for the left and right columns respectively. The vertical dotted brown line represents the time for which \(\tilde{k}_\perp = 1\), the point at which MHD and gyrokinetics are expected to diverge.
    }
    \label{numerical:fig2}
\end{figure}

Our flow shear results are shown in \autoref{numerical:fig2}.\footnote{We note to the reader that the significant divergences between GX and the analytical gyrokinetic solution in some of the simulations are attributed to the poor convergence of the Laguerre spectra for large \(k_\perp \rho_i\) for the linear gyrokinetic equation.} We first discuss the strong source regime in the left column to highlight how the gyrokinetic result is influenced by initial conditions. As expected, both gyrokinetics and MHD predict that an increased shear results in a suppression of the transient amplification, with their solutions diverging at \(\tilde{k}_\perp = 1\) as MHD inherently assumes that \(\tilde{k}_\perp \ll 1\). However, while the MHD solution FLR stabilises as expected from the zero shear result, the gyrokinetic solution becomes amplified. It is only much later that the FLR stabilisation sets in. As mentioned earlier in \S\ref{SECT: numerical - init cond}, this occurs despite the fact that we ensured both MHD and gyrokinetics would have the same initial condition and is due to the fact that FLR effects are embedded in the initial conditions for gyrokinetics and cannot be captured by MHD. What makes this physically interesting is that while we had set the initial condition to be a perturbed Maxwellian in \eqref{GK:eqn32}, the size of the perturbation, governed by the parameter \(\kappa\), does not appear in our equations. This implies that, regardless of the size of the perturbation, a small density gradient would result in a delayed FLR stabilisation.

\par 
On the other hand, for the weak source regime in the right column, where FLR physics dominates, we see that both gyrokinetics and MHD stabilise in a similar manner. Because the physically interesting regime occurs at \(\tilde{k}_\perp \ll 1\), we do not observe the explicit divergence at \(\tilde{k}_\perp\sim 1\) as we do in the strong source regime. In our opinion, we do not believe that this implies the MHD solution can be used in place of gyrokinetics. The reason is that MHD appears to consistently \textit{overestimate} the maximal amplification. It is likely that, as depicted in the figures, this is attributed to the fact that even at zero shear, there are small deviations between the gyrokinetic and MHD solutions, resolved only by taking the \(m_e/m_i \ll 1\) and \(\tilde{k}_y \ll 1\) limits of the former, details of which are described in \autoref{appx GK low order}.

\par
Another interesting physical result is the oscillatory behaviour of the solution after FLR stabilisation. We had initially believed this to be a symptom of the \(\tilde{k}_y \ll 1\) assumption. Surprisingly, this is also predicted in the gyrokinetic solution despite retaining all orders of \(\tilde{k}_y\). 

\subsubsection{Quantitative scaling of the transient amplification with shear}

\begin{figure}
    \centering
    \includegraphics[scale = 0.5]{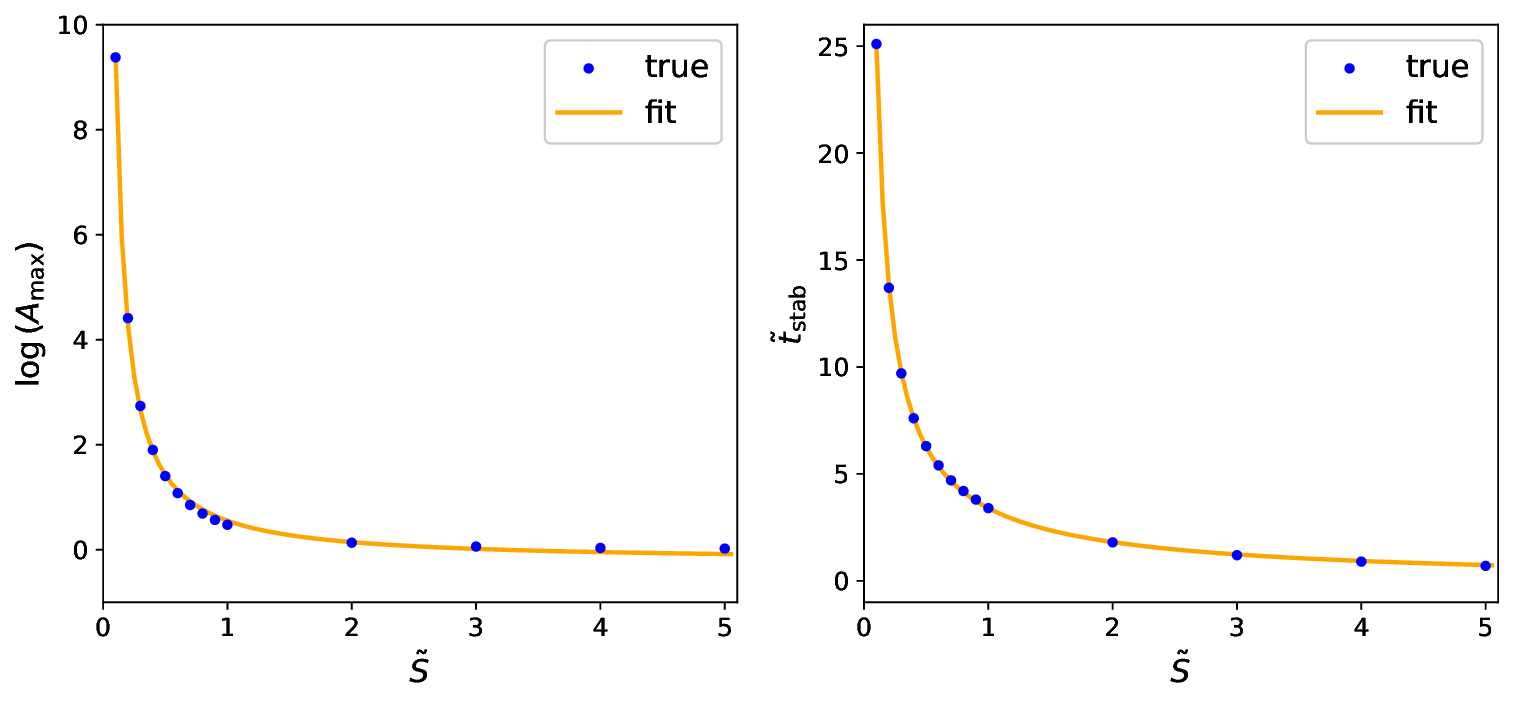}
    \captionsetup{width = 0.9\textwidth}
    \caption{
        Fit of the form \eqref{numerical:eqn13} for the logarithm of the maximal amplification \(\log\left(A_{\text{max}}\right)\) and the stabilisation time \(\tilde{t}_{\text{stab}}\) to the shear parameter \(\tilde{S}\)(orange line) compared to the analytical gyrokinetic solution (blue dots). 
    }
    \label{numerical:fig3}
\end{figure}

\begin{table}
    \centering
    \setlength{\tabcolsep}{8pt}
    \begin{tabular}{ c c c c c}
        
        &
        \(c_1\)
        &
        \(c_2\)
        &
        \(c_3\)
        &
        RMSE
        \\ \hline
        \(\log\left( A_{\text{max}} \right)\)
        &
        0.7654
        &
        1.1000
        &
        -0.2116
        &
        0.0706
        \\[0.5em]
        \(\tilde{t}_{\text{stab}}\)
        &
        3.5700
        &
        0.8490
        &
        -0.1694
        &
        0.0537
    \end{tabular}
    \captionsetup{width = 0.9\textwidth}
    \caption{Constants for the fit as well as the root mean squared error (RMSE).}
    \label{table2}
\end{table}

Focusing on the physically relevant weak source regime, let us denote the maximal transient amplification to be \(A_{\text{max}}\). This maximal amplification occurs at the stabilisation time, which we shall denote as \(\tilde{t}_\text{stab}\). For the range of \(0.1 \leq \tilde{S} \leq 5\), it is interesting to note that \(\log \left(A_{\text{max}}\right)\) and \(\tilde{t}\) can be fitted to \(\tilde{S}\) with the following relation:
\begin{equation}
    \begin{aligned}
        c_1 \tilde{S}^{-c_2} + c_3
        ,
    \end{aligned}
    \label{numerical:eqn13}
\end{equation}
where the constants \(c_1, c_2, c_3\) are given in \autoref{table2}. Based on the very small RMSE error in both cases, we see that the fit in \autoref{numerical:fig3} is excellent. Because \(c_2 > 0\) in both cases, the scaling relation seems to imply that as \(\tilde{S}\rightarrow +\infty\), no transient amplification will occur. In other words, at large shears, the instability would not have a chance to grow as it is immediately suppressed.

\section{Conclusion and future work} \label{SECT: conclusion}

This paper presents two models: one based on isothermal compressible MHD inspired by \citet{RobTay62} and \citet{Ng05}, and another based on gyrokinetics in an attempt to describe flow shear induced FLR stabilisation of flute interchanges in slab geometry. 

\par 
At zero shear, in contrast to ideal MHD, where all modes are predicted to be unstable, we demonstrated that by including FLR effects, which are inherent in gyrokinetics and achieved in MHD by adding gyroviscosity, there exists a threshold value of \(k_\perp \rho_i\) beyond which all modes are stable. The numerical results for this section also seem to demonstrate the interesting result that weaker instabilities take longer to FLR stabilise.

\par 
For a system with finite shear, an initially unstable mode with \(k_\perp\rho_i \ll 1\) will eventually become stable. This is because the wavenumbers are advected in time, eventually crossing the aforementioned threshold. However, in the time before stabilisation, transient amplification can still occur. Nevertheless, we were able to numerically demonstrate that the maximal transient amplification will still be suppressed with increasing shear. 

\par 
A key finding was that the gyrokinetic solution will be influenced by the initial condition in which its effect is controlled by the size of the density gradient rather than by that of the perturbation. This initial condition cannot be fully replicated in MHD as it has embedded FLR effects. As a result, this has led us to consider two regimes: the weak and strong source regimes. The former has a much smaller density gradient than the latter. 

\par 
In the strong source regime, the inherent FLR effects in the initial condition would result in a delayed FLR stabilisation compared to MHD. As the delay results in it occurring at \(k_\perp\rho_i \gg 1\), the MHD and gyrokinetic solutions diverge naturally at \(k_\perp\rho_i\sim1\), which is expected because of the long wavelength assumption in MHD.

\par 
In the weak source regime where the FLR physics is dominant, both the gyrokinetic and MHD solution predicts the same FLR stabilisation time, consistent with the result from our zero shear scenario. However, it was shown that MHD appears to consistently overestimate the maximal transient amplification.

\par 
We now discuss some possible extensions to the work at hand. To begin with, we note that we have focused mainly on electrostatic perturbations. As explained in the introduction, we have done so because it is the fluctuating electric field that is responsible for driving the instability. Regardless, it is entirely possible to include a perturbed magnetic field in this problem and consider full electromagnetic effects. 


\par 
We had also chosen to focus on flute modes (\(k_\parallel \rightarrow 0\)) as they are the most unstable. However, the above analysis can be extended to include finite \(k_\parallel\) effects in a manner similar to \citet{Alex12}. It would be interesting to compare the shear stabilisation of these modes with the flute modes.

\par 
Another feature that this work did not include is that of collisions. As already mentioned, we had neglected this because we are utilising previously derived collisionless models as our base of comparison. Despite this, we should note that the linearised collision operators used in gyrokinetics, contains terms that are proportional to \(k_\perp^2 \rho_i^2\) \citep{Abel08}. Hence, these terms may potentially be significant as \(k_\perp \rho_i \rightarrow 1\), making it valuable to understand their role in the problem in the future.

\section{Acknowledgements}

We express our gratitude to M. Landreman and the stellarator group at the University of Maryland, College Park, for the insightful discussions.

This research used resources from the National Energy Research Scientific Computing Centre (NERSC), a Department of Energy User Facility, using NERSC award FES-ERCAP 36155.

This material is based upon work supported by the U.S. Department of Energy, Office of Science, Office of Fusion Energy Sciences, through the SciDAC programme under Award Number DE-SC0024425 and Award Number DEFG0293ER54197. 

\section{Declaration of Interests}
Dr Ian Abel is the co-founder of Gridfire Inc. a company engaged in the commercialisation of the centrifugal mirror concept.

\appendix 

\section{Derivation of the extended MHD model with gyroviscosity} \label{appx MHD}

\subsection{Extended MHD equations}

The extended MHD equations found in \citet{RobTay62} comprises of the following equations:
\begin{align}
    &
    \text{Momentum: }
    m_i N_i \left(\frac{\partial}{\partial t} + \bsym{U} \cdot \nabla \right) \bsym{U} 
	=
	- \nabla (p_i + p_e) + m_i N_i \bsym{g} + \nu_i \bsym{\lambda}_i
    ,
    \label{appx MHD: eqn1}
    \\
    &
    \text{Continuity: }
    \frac{\partial N_i}{\partial t} + \nabla \cdot \left( N_i \boldsymbol{U} \right) = 0
    ,
    \label{appx MHD: eqn2}
    \\
    &
    \text{Ohm's law: }
    \boldsymbol{E} + \boldsymbol{U} \times \boldsymbol{B} + \frac{c}{eN_i} \nabla p_e
    =
    0
    ,
    \label{appx MHD: eqn3}
\end{align}
where we have neglected electron inertia and electron gyroviscosity. The capital letter quantities \(N_i, \bsym{U}_i\) consist of both the equilibrium quantity and its perturbation (\(N_i = n_i + \delta n_i\) and \(\bsym{U}_i = Sx\boldhat{y} + \delta\bsym{u}\)). We have elected not to capitalise the electron and ion pressure \(p_s\) because, as we shall soon see, the perturbed pressure \(\delta p\) will not appear in our equations explicitly. Also, we have \(\nu_i = \rho_i^2 \Omega_i/4\) and
\begin{align}
    &
    \lambda_x = 
    m_i
	\frac{\partial}{\partial x} \left[ N_i \left(\frac{\partial U_y}{\partial x} + \frac{\partial U_x}{\partial y} \right) \right]
	-
    m_i
	\frac{\partial}{\partial y} \left[ N_i \left(\frac{\partial U_x}{\partial x} - \frac{\partial U_y}{\partial y} \right) \right]
    ,
	\label{appx MHD: eqn4}
	\\
	&
	\lambda_y = 
	-m_i \frac{\partial}{\partial y} \left[ N_i \left(\frac{\partial U_y}{\partial x} + \frac{\partial U_x}{\partial y} \right) \right]
	-
    m_i 
	\frac{\partial}{\partial x} \left[ N_i \left(\frac{\partial U_x}{\partial x} - \frac{\partial U_y}{\partial y} \right) \right]
    ,
	\label{appx MHD: eqn5}
\end{align}
where the quantity \(\bsym{\lambda}\) is simply equivalent to \(-\nabla\cdot\boldsymbol{\Pi}_{i\times}\) where \(\boldsymbol{\Pi}_{i\times}\) is the gyro-viscosity component of the Braginskii stress tensor \citep{Braginskii65, Schnack06, parra19a}. We will also assume our MHD model to have isothermal i.e. \(p_s = n_s T_s\) with \(T_s = \text{constant}\), as was done in \citet{Ng05}.

\par 
It is convenient to substitute the momentum equation \eqref{appx MHD: eqn1} as an expression for \(p_e\) in \eqref{appx MHD: eqn3}. This gives us
\begin{equation}
    \begin{aligned}
        & 
        \boldsymbol{E} + \boldsymbol{U} \times \boldsymbol{B} 
		- \frac{c}{en} \nabla p_i - \frac{m_i c}{e}\left(\frac{\partial}{\partial t} + \boldsymbol{U} \cdot \nabla \right) \boldsymbol{U}  + \frac{m_i c}{e} \boldsymbol{g} + \frac{c}{e}\frac{\nu_i }{N_i} \boldsymbol{\lambda}
		=
		0
        .
    \end{aligned}
    \label{appx MHD: eqn6}
\end{equation}
Taking the curl of this equation and, by assuming a constant magnetic field \(\bm{B} = B \boldhat{z}\), we can utilise Faraday's law \(c \nabla\times\boldsymbol{E} = \partial_t\boldsymbol{B} = 0\) to give us
\begin{equation}
    \begin{aligned}
        & 
        \boldhat{z} \nabla\cdot\bsym{U} + \frac{1}{\Omega_i} \nabla\times \left(\frac{\partial}{\partial t} + \bsym{U} \cdot \nabla \right) \bsym{U}  
        -\frac{\nu_i }{m_i \Omega_i} \nabla\times \left( \frac{1}{ N_i} \boldsymbol{\lambda}_i \right)
        =
        0
        .
    \end{aligned}
    \label{appx MHD: eqn7}
\end{equation}
Equations \eqref{appx MHD: eqn1}, \eqref{appx MHD: eqn2}, and \eqref{appx MHD: eqn7} are our extended MHD equations

\subsection{Linearisation of extended MHD equations} 

\subsubsection{Ordering assumptions}

We want to linearise the extended MHD equations \eqref{appx MHD: eqn1}, \eqref{appx MHD: eqn2}, and \eqref{appx MHD: eqn7}. However, because the linearised equations are too complicated for our purposes, we will take inspiration from the gyrokinetic ordering \citep{Abel13} to neglect certain small terms. We shall assume that
\begin{equation}
    \begin{aligned}
        & \frac{\partial}{\partial t}  \sim \omega \sim \epsilon \Omega_i,
        \qquad \frac{\partial}{\partial x} \sim \frac{\partial}{\partial y} \sim k_\perp \sim \frac{1}{\rho_i},
        \qquad 
        \frac{\rmd \log n_i}{\rmd x} \sim \frac{1}{a}
        .
    \end{aligned}
    \label{appx MHD: eqn8}
\end{equation}
For now, we will still assume \(k_\perp \rho_i \sim 1\) and it is only later onwards will we take \(k_\perp \rho_i \ll 1\) by considering a subsidiary expansion in small \(k_\perp \rho_i\) \citep{Alex09}.

\par 
We need to come up with the ordering assumptions for \(S, \delta u_x, \delta u_y\), and \(\delta n_i\). We assume \(\delta n_i \sim \epsilon n_i\) and \(\delta u_x \sim \delta u_y \sim \epsilon \vths{i}\). For \(S\), we are interested in the regime where
\begin{equation}
    \begin{aligned}
        S \sim \gamma_g \sim \frac{\vths{i}}{a} \sim \omega
        ,
    \end{aligned}
    \label{appx MHD: eqn9}
\end{equation}
where \(\gamma_g\) is the ideal interchange growth rate given in \eqref{MHD:eqn1}. 

\subsubsection{Linearised equations}

Firstly, for the momentum equation \eqref{appx MHD: eqn1}, we take the curl, linearise it, and divide throughout by \(n_i\). In the flute limit \(k_\parallel \rightarrow 0\), we obtain
\begin{equation}
    \begin{aligned}[b]
        & \underbrace{\frac{\partial}{\partial t}}_{\sim \epsilon \Omega_i} \Bigg[ 
             \underbrace{\left( 
                \frac{\partial \delta u_y}{\partial x}
                - \frac{\partial \delta u_x}{\partial y}
            \right)}_{\sim \epsilon \Omega_i }
            + 
            \underbrace{\frac{\rmd \log n_i}{\rmd x} \delta u_y}_{\sim \epsilon^2 \Omega_i}
        \Bigg]
        +
        \underbrace{S}_{\epsilon \Omega_i}
        \Bigg[
            \underbrace{\frac{\rmd \log n_i}{\rmd x} \left( \delta u_x 
                + x \frac{\partial \delta u_y}{\partial y} 
            \right)}_{\sim \epsilon^2 \Omega_i }
        \\
        &
            +  
                \underbrace{\frac{\partial \delta u_x}{\partial x}
                +  \frac{\partial \delta u_y}{\partial y}
                + x \frac{\partial^2 \delta u_y}{\partial x \partial y}
                - x \frac{\partial^2 \delta u_x}{\partial y^2}
                    }_{\sim \epsilon \Omega_i }
        \Bigg]
        \\
        &=
        \underbrace{g \frac{\partial}{\partial y} \left(\frac{\delta n_i}{n_i} \right)}_{\sim \epsilon^2 \Omega_i^2 }
        +
        \underbrace{\nu_i}_{\sim \epsilon^2 a^2 \Omega_i} \Bigg\{
            -
            \underbrace{ 
            \frac{\rmd^2 \log n_i}{\rmd x^2} \left( \frac{\partial \delta u_x}{\partial x} - \frac{\partial \delta u_y}{\partial y} \right) 
                }_{\sim \epsilon \Omega_i/a^2 }
        \\ 
        &
            -
            \underbrace{2 \frac{\rmd \log n_i}{\rmd x} \left( 
                \frac{\partial^2 \delta u_x}{\partial x^2}
                +
                \frac{\partial^2 \delta u_x}{\partial y^2}
            \right) }_{\sim \Omega_i /a^2}
            \underbrace{
            - 
                \frac{\partial^3 \delta u_y}{\partial y \partial x^2}
                -\frac{\partial^3 \delta u_x}{\partial y^2 \partial x} 
            -
              \frac{\partial^3 \delta u_x}{\partial x^3} -
             \frac{\partial^3 \delta u_y}{\partial y^3}
                }_{\sim \Omega_i / \epsilon a^2}
            -
            \underbrace{
            2 S \frac{\partial^2 }{\partial x \partial y}
            \left( \frac{\delta n_i}{n_i} \right)
                }_{\sim \Omega_i /a^2}
         \Bigg\} 
         .
    \end{aligned}
    \label{appx MHD: eqn10}
\end{equation}
We only retain terms that are comparable in size to the lowest non-vanishing time-derivative on the LHS; therefore, we retain only up to \(\mathcal{O}(\epsilon^2)\) and the equation reduces to
\begin{equation}
    \begin{aligned}
        &\left(\frac{\partial}{\partial t}  + S x \frac{\partial}{\partial y} \right)
             \left( 
                \frac{\partial \delta u_y}{\partial x}
                - \frac{\partial \delta u_x}{\partial y}
            \right)
        +
        S \left( 
                \frac{\partial \delta u_x}{\partial x}
                +  \frac{\partial \delta u_y}{\partial y}
        \right)
       =
        g \frac{\partial}{\partial y} \left(\frac{\delta n_i}{n_i} \right)
       \\
        &+
        \nu_i \bigg[
            -
            2 \frac{\rmd \log n_i }{\rmd x} \left( 
                \frac{\partial^2 \delta u_x}{\partial x^2}
                +
                \frac{\partial^2 \delta u_x}{\partial y^2}
            \right) 
            - 
                \frac{\partial^3 \delta u_y}{\partial y \partial x^2}
                -\frac{\partial^3 \delta u_x}{\partial y^2 \partial x}
            -
             \frac{\partial^3 \delta u_x}{\partial x^3} -
             \frac{\partial^3 \delta u_y}{\partial y^3}
         \\
         &
            -
            2 S \frac{\partial^2 }{\partial x \partial y}
            \left( \frac{\delta n_i}{n_i} \right)
         \bigg]
         .
    \end{aligned}
    \label{appx MHD: eqn11}
\end{equation}

\par 
Next, for the continuity equation \eqref{appx MHD: eqn2}, we linearise and divide throughout by \(n_i\):
\begin{equation}
    \begin{aligned}
        \left( 
        \frac{\partial }{\partial t} 
        + S  x \frac{\partial}{\partial y }
        \right) \frac{\delta n_i}{n_i}
        +
        \left(
            \frac{\partial \delta u_x}{\partial x}
            + \frac{\partial \delta u_y}{\partial y}
        \right) 
        +
        \frac{\rmd \log n_i}{\rmd x} \delta u_x
        = 0
        .
    \end{aligned}
    \label{appx MHD: eqn12}
\end{equation}
No ordering approximations are required here since all terms are \(\mathcal{O}(\epsilon)\) or larger.

\par 
Lastly, for Ohm's law \eqref{appx MHD: eqn7}, we take the dot product with \(\boldhat{z}\) and linearise:
\begin{equation}
    \begin{aligned}
        &
        \underbrace{
        \frac{\partial \delta u_x}{\partial x} 
        + \frac{\partial \delta u_y}{\partial y}
            }_{\sim \epsilon \Omega_i }
        +
        \frac{1}{\Omega_i} 
        \Bigg[
            \underbrace{
            \left(\frac{\partial}{\partial t} + S x \frac{\partial}{\partial y} \right)
            }_{\sim \epsilon \Omega_i}
            \underbrace{
            \left(
                \frac{\partial \delta u_y}{\partial x}
                - \frac{\partial \delta u_x}{\partial y}
            \right) 
            }_{\sim \epsilon \Omega_i }
            +
            \underbrace{
            S \left(
                \frac{\partial \delta u_x}{\partial x}
                + \frac{\partial \delta u_y}{\partial y}
            \right) 
                }_{\sim \epsilon^2 \Omega_i^2 }
        \Bigg]
        \\
        &
        +
        \underbrace{\frac{\nu_i}{\Omega_i}}_{\sim \epsilon^2 a^2}
        \left\{
            \begin{aligned}
                &
                \underbrace{
                 \frac{\rmd^2 \log n_i}{\rmd x^2} \left( \frac{\partial \delta u_x}{\partial x} - \frac{\partial \delta u_y}{\partial y} \right) 
                -
                \left(\frac{\rmd \log n_i}{\rmd x} \right)^2 \left(\frac{\partial \delta u_x}{\partial x} - \frac{\partial \delta u_y}{\partial y} \right)
                    }_{\sim \epsilon \Omega_i /a^2 }
                 \\
                 &
               +
                \underbrace{
                \frac{\rmd \log n_i}{\rmd x}\left(\frac{\partial^2 \delta u_x}{\partial x^2}  
                + \frac{\partial^2 \delta u_x}{\partial y^2} \right)
                    }_{\sim \Omega_i /a^2}
                 -
                 \underbrace{
                 2 S
                \frac{\rmd \log n_i}{\rmd x}
                \frac{\partial }{\partial y} \left(\frac{\delta n_i}{n_i} \right)
                    }_{\sim \epsilon \Omega_i / a^2}
                \\
                &
                +
                \underbrace{
                    \frac{\partial^3 \delta u_y}{\partial y \partial x^2}
                    + \frac{\partial^3 \delta u_x}{\partial y^2 \partial x}
                +
                 \frac{\partial^3 \delta u_x}{\partial x^3} + \frac{\partial^3 \delta u_y}{\partial y^3} 
                    }_{\sim \Omega_i / \epsilon a^2}
                +
                \underbrace{
                2 S \frac{\partial^2 }{\partial x \partial y} \left(\frac{\delta n_i}{n_i} \right)
                    }
                    _{\sim \Omega_i /a^2}
            \end{aligned}
        \right\}
        \\
        &
        = 0
        .
    \end{aligned}
    \label{appx MHD: eqn13}
\end{equation}
Again, retaining up to only \(\mathcal{O}(\epsilon^2)\), we obtain
\begin{equation}
    \begin{aligned}
        &
        \frac{\partial \delta u_x}{\partial x} 
        + \frac{\partial \delta u_y}{\partial y}
        +
        \frac{1}{\Omega_i} 
        \left[
            \left(\frac{\partial}{\partial t} + S x \frac{\partial}{\partial y} \right)
            \left(
                \frac{\partial \delta u_y}{\partial x}
                - \frac{\partial \delta u_x}{\partial y}
            \right) 
            +
            S \left(
                \frac{\partial \delta u_x}{\partial x}
                + \frac{\partial \delta u_y}{\partial y}
            \right) 
        \right]
        \\
        &
        +
        \frac{\nu_i}{\Omega_i} 
        \left[ 
            \frac{\rmd \log n_i}{\rmd x}\left(\frac{\partial^2 \delta u_x}{\partial x^2}  
                + \frac{\partial^2 \delta u_x}{\partial y^2} \right)
            +
                    \frac{\partial^3 \delta u_y}{\partial y \partial x^2}
                    +\frac{\partial^3 \delta u_x}{\partial y^2 \partial x}
                +
                 \frac{\partial^3 \delta u_x}{\partial x^3} + \frac{\partial^3 \delta u_y}{\partial y^3} 
        \right. 
        \\
        &\hspace{1.2cm} \left.
            +
            2 S \frac{\partial^2 }{\partial x \partial y} \left(\frac{\delta n_i}{n_i} \right)
        \right]
        = 0
        .
    \end{aligned}
    \label{appx MHD: eqn14}
\end{equation}
The three equations \eqref{appx MHD: eqn11}, \eqref{appx MHD: eqn12}, and \eqref{appx MHD: eqn14} are the linearised equations that solve for the three unknowns \(\delta n_i\), \(\delta u_x\), and \(\delta u_y\).

\subsection{Zero shear dispersion relation} \label{appx MHD SECT: zero shear}

In the case of zero shear (\(S = 0\)), the three equations become
\begin{align}
    &
    \begin{aligned}[b]
    &\frac{\partial}{\partial t} 
     \left( 
        \frac{\partial \delta u_y}{\partial x}
        - \frac{\partial \delta u_x}{\partial y}
     \right)
    =
    g \frac{\partial}{\partial y} \left(\frac{\delta n_i}{n_i} \right)
    +
    \nu_i \bigg[
        -
        2 \frac{\rmd \log n_i}{\rmd x} \left( 
            \frac{\partial^2 \delta u_x}{\partial x^2}
            +
            \frac{\partial^2 \delta u_x}{\partial y^2}
        \right) 
    \\
    &- 
            \frac{\partial^3 \delta u_y}{\partial y \partial x^2}
            -\frac{\partial^3 \delta u_x}{\partial y^2 \partial x}
        -
         \frac{\partial^3 \delta u_x}{\partial x^3} -
         \frac{\partial^3 \delta u_y}{\partial y^3}
     \bigg]
     ,
     \end{aligned}
    \label{appx MHD: eqn15}
    \\[0.5em]
    &
    \frac{\partial }{\partial t} \frac{\delta n_i}{n_i}
    +
    \left(
        \frac{\partial \delta u_x}{\partial x}
        + \frac{\partial \delta u_y}{\partial y}
    \right) 
    +
    \frac{\rmd \log n_i}{\rmd x} \delta u_x
    = 0
    ,
    \label{appx MHD: eqn16}
    \\[0.5em]
    &
    \begin{aligned}[b]
        &
        \frac{\partial \delta u_x}{\partial x} 
        + 
        \frac{\partial \delta u_y}{\partial y}
        +
        \frac{1}{\Omega_i} 
        \frac{\partial}{\partial t} 
        \left(
            \frac{\partial \delta u_y}{\partial x}
            - \frac{\partial \delta u_x}{\partial y}
        \right) 
        +
        \frac{\nu_i}{\Omega_i} 
        \bigg[ 
            \frac{\rmd \log n_i}{\rmd x}\left(\frac{\partial^2 \delta u_x}{\partial x^2} 
                + 
                \frac{\partial^2 \delta u_x}{\partial y^2} 
                \right)
        \\
        & 
        +
        \frac{\partial^3 \delta u_y}{\partial y \partial x^2}
        +
        \frac{\partial^3 \delta u_x}{\partial y^2 \partial x}
        +
        \frac{\partial^3 \delta u_x}{\partial x^3} 
        + 
        \frac{\partial^3 \delta u_y}{\partial y^3} 
        \bigg]
        = 0
        .
    \end{aligned}
    \label{appx MHD: eqn17}
\end{align}
As we expect time-independent growth rates, we can assume normal mode solutions. The perturbed quantities can then be written as:
\begin{equation}
        X(\boldsymbol{r}, t) 
        =
        \hat{X} e^{\im k_y y + \im k_x x - \im\omega t}
        .
    \label{appx MHD: eqn18}
\end{equation}
Substituting this into \eqref{appx MHD: eqn15}--\eqref{appx MHD: eqn17}, we obtain
\begingroup 
\allowdisplaybreaks
\begin{align}
    &
    \begin{aligned}[b]
        &  \omega
         \left( 
             k_x \widehat{\delta u}_y
            -  k_y \widehat{\delta u}_x
         \right)
        =
        \im k_y  g  \frac{\widehat{\delta n}_i}{n_i} 
        +
        \nu_i \bigg[
            2 \frac{\rmd \log n_i}{\rmd x} \left( 
                k_x^2
                +
    			 k_y^2
            \right)  \widehat{\delta u}_x
        \\
        &+
    	\im k_y k_x^2 \widehat{\delta u}_y
    	+
    	\im k_y^2 k_x \widehat{\delta u}_x
    	+
    	\im k_x^3 \widehat{\delta u}_x
    	+
    	\im k_y^3 \widehat{\delta u}_y
         \bigg]
         ,
     \end{aligned}
    \label{appx MHD: eqn19}
    \\[0.5em]
    &
    -\im\omega \frac{\widehat{\delta n}_i}{n_i}
    +
    \left[
    	(\im k_x) \widehat{\delta u}_x
    	+ 
    	(\im k_y) \widehat{\delta u}_y
    \right]
    +
    \frac{\rmd \log n_i}{\rmd x} \widehat{\delta u}_x
    = 0
    ,
    \label{appx MHD: eqn20}
    \\[0.5em]
    &
    \begin{aligned}[b]
    	&
    	\im k_x \widehat{\delta u}_x 
    	+ 
    	\im k_y \widehat{\delta u}_y
    	+
    	\frac{\omega}{\Omega_i}  
    	\left(
    		k_x \widehat{\delta u}_y 
    		-
    		 k_y \widehat{\delta u}_x
    	\right)
    	+
    	\frac{\nu_i}{\Omega_i} 
    	\bigg[ 
    		-
    		\frac{\rmd \log n_i}{\rmd x}
    		\left(k_x^2 + k_y^2 \right) \widehat{\delta u}_x
    	\\
    	& 
    	-
    	\im k_x^2  k_y \widehat{\delta u}_y
    	-
    	\im k_x k_y^2 \widehat{\delta u}_x
    	-
    	\im k_x^3 \widehat{\delta u}_x
    	-
    	\im k_y^3 \widehat{\delta u}_y
    	\bigg]
    	= 0
        .
    \end{aligned}
    \label{appx MHD: eqn21}
\end{align}
\endgroup 
This system of three equations can be solved for the three unknowns \(\widehat{\delta u}_x\), \(\widehat{\delta u}_y\), and \(\widehat{\delta n}_i\), giving us the following dispersion relation:
\begin{equation}
    \begin{aligned}
	&
	\bigg(
		\underbrace{\frac{k_x^2}{k_y^2}}_{\sim 1}
		-
		\underbrace{\im\frac{\rho_i^2}{4} \frac{\rmd \log n_i}{\rmd x} 
		k_x
		\frac{ k_\perp^2}{k_y^2}}_{\sim \epsilon}
		+
		1
	\bigg)
	\underbrace{\omega^2}_{\sim \epsilon^2 \Omega_i^2}
	+
	\bigg[
		-
		\underbrace{(\omega_{gi} + \omega_{*i}) \frac{k_\perp^2}{k_y^2}}_{\sim \epsilon \Omega_i}
		+
		\underbrace{\im \frac{\rmd \log n_i}{\rmd x} \omega_{gi} \frac{k_x}{k_y^2} }_{\sim \epsilon^2 \Omega_i}
    \\
    &+
		\underbrace{
		\frac{1}{8}\omega_{*i} \frac{k_\perp^4}{k_y^2} \rho_i^2
		}_{\sim \epsilon \Omega_i}
	\bigg]
	\underbrace{\omega}_{\sim \epsilon \Omega_i}
	=
	- 
	\underbrace{
	\frac{\omega_{gi}\omega_{*i}}{k_y^2 \rho_i^2/2}\
		}_{\sim \epsilon^2 \Omega_i^2}
    .
    \end{aligned}
    \label{appx MHD: eqn22}
\end{equation}
As before, we retain only up to \(\mathcal{O}(\epsilon^2)\). The lowest order i.e. \(\mathcal{O}(1)\) gives us
\begin{equation}
    \left(\frac{k_x^2}{k_y^2}+1\right)\omega^2
    =
    - \frac{\omega_{gi}\omega_{*i}}{k_y^2 \rho_i^2/2}
    \Longrightarrow
    \omega^2 
    =
    - \frac{\omega_{gi}\omega_{*i}}{k_\perp^2 \rho_i^2/2}
    .
    \label{appx MHD: eqn23}
\end{equation}
In other words, equation \eqref{appx MHD: eqn23} tells us that any finite value of \(k_x\) contributes to the stability of the mode. As such, we will set \(k_x = 0\) and hence \(k_\perp = k_y\) in order to focus on the most unstable modes. Then, retaining only up to \(\mathcal{O}(\epsilon^2)\) in \eqref{appx MHD: eqn22} gives us \eqref{MHD:eqn4}.

\subsection{Time-dependent finite shear dispersion relation} \label{appx MHD SECT: time dep}

From \eqref{appx MHD: eqn11}, \eqref{appx MHD: eqn12}, and \eqref{appx MHD: eqn14} in which \(S \neq 0\), we can apply the coordinate transformation \eqref{MHD:eqn7} to obtain:
\begingroup
\allowdisplaybreaks
\begin{align}
    &
    \begin{aligned}[b]
        &
    	-
    	St'\frac{\partial}{\partial t'}  \left( \frac{\partial \delta u_y}{\partial y'}\right)
    	-
    	\frac{\partial}{\partial t'}  \frac{\partial \delta u_x}{\partial y'}
    	-
    	S^2 t' \frac{\partial \delta u_x}{\partial y'}
        =
    	g \frac{\partial}{\partial y'} \left(\frac{\delta n_i}{n_i} \right)
    	\\[0.5em]
    	& 
    	+
    	\nu_i \left(S^2 {t'}^2 +1\right)\left[
		-
			2 \frac{\rmd \log n_i}{\rmd x} 
			\frac{\partial^2 \delta u_x}{\partial {y'}^2}
		+ 
		S {t'} 
    		 \frac{\partial^3 \delta u_x}{\partial {y'}^3}
    		 -
    		 \frac{\partial^3 \delta u_y}{\partial {y'}^3}
    		 \right]
    		+  2\nu_i 
    		 S^2 t' \frac{\partial^2 }{\partial {y'}^2}
		\left( \frac{\delta n_i}{n_i} \right)
        ,
    \end{aligned}
     \label{appx MHD: eqn24}
    \\[0.5em]
    &
    \begin{aligned}
        \frac{\partial }{\partial t'} \left(\frac{\delta n_i}{n_i} \right)
        +
        \left(
        	- S t' \frac{\partial \delta u_x}{\partial y'}
        	+ \frac{\partial \delta u_y}{\partial y'}
        \right) 
        +
        \frac{\rmd \log n_i}{\rmd x} \delta u_x
        = 0
        ,
    \end{aligned}
     \label{appx MHD: eqn25}
    \\[0.5em]
    &
    \begin{aligned}[b]
        &
        - S t'\frac{\partial \delta u_x}{\partial y'} 
        + \frac{\partial \delta u_y}{\partial y'}
        +
        \frac{1}{\Omega_i} 
        \left[
            - St' \frac{\partial}{\partial t'}  \left( \frac{\partial \delta u_y}{\partial y'} \right)
		-
		\frac{\partial}{\partial t'} \left(\frac{\partial \delta u_x}{\partial y'}\right)
		-
		S^2 t' \frac{\partial \delta u_x}{\partial y'}
        \right]
        \\
        &
        +
        \frac{\nu_i}{\Omega_i} \left(
			S^2 {t'}^2 + 1
                \right)
        \left[ 
            \frac{\rmd \log n_i}{\rmd x}\frac{\partial^2 \delta u_x}{\partial {y'}^2}
		-
		S t'
		 \frac{\partial^3 \delta u_x}{\partial {y'}^3} 
            +
                    \frac{\partial^3 \delta u_y}{\partial {y'}^3}
        \right]
		-
            \frac{2\nu_i}{\Omega_i} S^2 t' \frac{\partial^2 }{\partial {y'}^2} \left(\frac{\delta n_i}{n_i} \right)
        = 0
        .
    \end{aligned}
    \label{appx MHD: eqn26}
\end{align}
\endgroup
We assume perturbed quantities of the form \(X = \hat{X}(t') e^{\im k_y' y'}\) where the quantity \(X\) refers to either \(\delta u_x, \delta u_y\), or \(\delta n_i\). After suppressing the primes in the above equations, we obtain:
\begin{align}
    &
    \begin{aligned}[b]
        &
        -
        \im St\frac{\partial \widehat{\delta u}_y}{\partial t}  
        -
        \im  \frac{\partial \widehat{\delta u}_x}{\partial t}
        -
        \im  S^2 t \widehat{\delta u}_x
        =
        \im  g \frac{\widehat{\delta n}_i}{n_i} 
        \\
        &
        +
        {k_y} \nu_i \left(S^2 {t}^2 +1\right)\left(
        2  \frac{\rmd \log n_i}{\rmd x}  \widehat{\delta u}_x
        -
        \im S {t} {k_y}  \widehat{\delta u}_x
        +
        \im{k_y} \widehat{\delta u}_y
        \right)
        -
        2\nu_i S^2 t {k_y}
        \frac{\widehat{\delta n}_i}{n_i}
        ,
    \end{aligned}
    \label{appx MHD: eqn27}
    \\[0.5em]
    &
    \begin{aligned}
        \frac{\partial }{\partial t} \left(\frac{\widehat{\delta n}_i}{n_i} \right)
	       - \im S t k_y \widehat{\delta u}_x
	       + \im k_y \widehat{\delta u}_y
        +
        \frac{\rmd \log n_i}{\rmd x} \widehat{\delta u}_x
        = 0
        ,
    \end{aligned}
    \label{appx MHD: eqn28}
    \\[0.5em]
    &
    \begin{aligned}[b]
        &
    	- \im S t\widehat{\delta u}_x
    	+ \im\widehat{\delta u}_y
    	-
    	\frac{\im }{\Omega_i} 
    	\left(
    	       St \frac{\partial \widehat{\delta u}_y }{\partial t}  
            +
           \frac{\partial  \widehat{\delta u}_x}{\partial t}
    	    +
    	       S^2 t \widehat{\delta u}_x
    	\right)
    	\\
    	&
    	+
    	\frac{{k_y} \nu_i}{\Omega_i} \left(
    		S^2 {t}^2 + 1
    			\right)
    	\left(
    		-\frac{\rmd \log n_i}{\rmd x}  \widehat{\delta u}_x
    	+
    	\im S t {k_y} \widehat{\delta u}_x
    	-
    	\im{k_y} \widehat{\delta u}_y
    	\right)
    	+
    	\frac{2\nu_i}{\Omega_i} S^2 t {k_y} \frac{\widehat{\delta n}_i}{n_i} 
    	= 0
        .
    \end{aligned}
    \label{appx MHD: eqn29}
\end{align}
Equation \eqref{appx MHD: eqn29} can be multiplied by \(\Omega_i\) and subtracted from \eqref{appx MHD: eqn27}. The resulting equation can then be combined with \eqref{appx MHD: eqn28} to give:
\begin{align}
    &
    \widehat{\delta u}_x 
    =
    \left(\frac{\rmd \log n_i}{\rmd x} \right)^{-1} 
    \frac{
    	\frac{\partial }{\partial t} \left(\frac{\widehat{\delta n}_i}{n_i} \right)
    	- \frac{\im g k_y }{\Omega_i} \frac{\widehat{\delta n}_i}{n_i} 
    }{
    	 \frac{{k_y}^2 \nu_i}{\Omega_i} \left(S^2 {t}^2 +1\right) - 1
    }
    ,
    \label{appx MHD: eqn30}
    \\
    &
    \widehat{\delta u}_y
    =
    -  \frac{g}{\Omega_i} \frac{\widehat{\delta n}_i}{n_i} 
    +
    \left[
    	\im \frac{\rmd \log n_i}{\rmd x}   \frac{{k_y} \nu_i}{\Omega_i} \left(S^2 {t}^2 +1\right)
    	+ S t
    \right]
    \left(\frac{\rmd \log n_i}{\rmd x} \right)^{-1} 
    \frac{
    	\frac{\partial }{\partial t} \left(\frac{\widehat{\delta n}_i}{n_i} \right)
    	- \frac{\im g k_y}{\Omega_i} \frac{\widehat{\delta n}_i}{n_i} 
    }{
    	 \frac{{k_y}^2 \nu_i}{\Omega_i} \left(S^2 {t}^2 +1\right) - 1
    }
    .
    \label{appx MHD: eqn31}
\end{align}
Substituting into \eqref{appx MHD: eqn27} and using the definitions of \(\omega_{*i}\) and \(\omega_{gi}\) in \eqref{MHD:eqn2} and \eqref{MHD:eqn3} respectively, gives:
\begin{equation}
    \begin{aligned}
        &
        0 = (S^2 t^2 + 1)
    	\bigg(
    			 \frac{\im}{2} \underbrace{\frac{\omega_{*i}}{\Omega_i}}_{\sim\epsilon}  S t + 1
    	\bigg)
    	\underbrace{\frac{\partial^2 }{\partial t^2} \left(\frac{\widehat{\delta n}_i}{n_i} \right)}_{\sim \epsilon^3 \Omega_i^2}
        \\
        &+
        \left\{
        \begin{aligned}
        	& 
        	\underbrace{St \frac{1}{\Omega_i} \frac{\omega_{*i}\omega_{gi}}{k_y^2 \rho_i^2/2}}_{\sim \epsilon^2 \Omega_i}
        	+
        	  \underbrace{\im S^3 t^2 \frac{\omega_{*i}}{\Omega_i}}_{\sim \epsilon^2 \Omega_i}
        	+
        	\underbrace{2 S^2 t }_{\sim \epsilon \Omega_i}
        	  -
        	  \underbrace{\im (\omega_{gi} + \omega_{*i})(S^2 t^2 + 1)}_{\sim \epsilon \Omega_i}
        	  \\
        	  &
        	+ 
        	 \underbrace{\im \frac{k_y^2 \rho_i^2 }{8} \omega_{*i} (S^2 t^2 + 1)^2}_{\sim\epsilon\Omega_i}
        	 -
        	 \underbrace{\frac{k_y^2 \rho_i^2}{2} 
        	\frac{
        		1 + \frac{\im St }{2}  \frac{\omega_{*i}}{\Omega_i}
        	}{
        		\frac{k_y^2 \rho_i^2 }{4} \left(S^2 t^2 + 1\right) - 1
        	}
        	S^2 t \left(S^2 t^2 +  1\right)
        		}_{\sim\epsilon\Omega_i}
        \end{aligned}
        \right\}
        \underbrace{\frac{\partial}{\partial t} \left(\frac{\widehat{\delta n}_i}{n_i}\right)}_{\sim \epsilon^2 \Omega_i}
        \\
        &
        +
        \left\{
        \begin{aligned}
        	&
        	\underbrace{\frac{\omega_{gi}\omega_{*i} }{\Omega_i} S^3 t^2}_{\sim \epsilon^3 \Omega_i^2}
        	-
        	 \underbrace{\im (2\omega_{gi} + \omega_{*i} ) S^2 t}_{\sim \epsilon^2 \Omega_i^2}
        	  +
        	 \underbrace{ 2 \omega_{*i} \omega_{gi} (S^2 t^2 + 1)}_{\sim \epsilon^2 \Omega_i^2}
              -
        	 \underbrace{\frac{\omega_{gi} \omega_{*i}}{k_y^2\rho_i^2/2} }_{\sim \epsilon^2 \Omega_i^2}
        	\\
        	&
        	+
        	\underbrace{\im \frac{k_y^2 \rho_i^2}{2}  
        	\frac{
        		1 + \frac{\im\omega_{*i}}{2 \Omega_i}St
        	}{
        	\frac{k_y^2 \rho_i^2}{4} (S^2 t^2 + 1) - 1
        	}
        	\omega_{gi} S^2 t(S^2 t^2 + 1)}_{\sim\epsilon^2 \Omega_i^2}
            +
             \underbrace{\im  \frac{k_y^2 \rho_i^2}{4} 
        	  \omega_{*i} S^2 t (S^2 t^2 + 1)}_{\sim \epsilon^2 \Omega_i^2}
        \end{aligned}
        \right\}
        \underbrace{\frac{\widehat{\delta n}_i }{n_i}}_{\sim \epsilon}
        .
    \end{aligned}
    \label{appx MHD: eqn32}
\end{equation}
We retain terms that are only order \(\mathcal{O}(\epsilon^3)\) since this is the lowest non-vanishing order of \(\partial^2 / \partial t^2\) on the LHS. This yields:
\begin{equation}
    \begin{aligned}[b]
        &
        (S^2 t^2 + 1) \frac{\partial^2 \widehat{\delta n}_i}{\partial t^2} 
        +
        \Bigg\{
            2 S^2 t 
            - 
            \im (\omega_{gi} + \omega_{*i})(S^2 t^2 + 1)
            + 
            \im \frac{k_y^2 \rho_i^2 }{8} \omega_{*i} (S^2 t^2 + 1)^2
        \\
        &
            -
            \frac{k_y^2 \rho_i^2}{2} 
        	\frac{S^2 t \left(S^2 t^2 +  1\right)}{\frac{k_y^2 \rho_i^2 }{4} \left(S^2 t^2 + 1\right) - 1}
        \Bigg\}\frac{\partial \widehat{\delta n}_i}{\partial t} 
        +
        \Bigg\{
        	-\im \left(2\omega_{gi} + \omega_{*i}\right) S^2 t
             +
             \im \frac{k_y^2 \rho_i^2}{2}  
        	\frac{
        		\omega_{gi} S^2 t(S^2 t^2 + 1)
        	}{
        	\frac{k_y^2 \rho_i^2}{4} (S^2 t^2 + 1) - 1
        	}
        \\
        &
        	+
            \im  \frac{k_y^2 \rho_i^2}{4} 
        	  \omega_{*i} S^2 t (S^2 t^2 + 1)
        	  +
        	  2 \omega_{*i} \omega_{gi} (S^2 t^2 + 1)  
            -
        	  \frac{\omega_{gi} \omega_{*i}}{k_y^2\rho_i^2/2} 
        \Bigg\}\widehat{\delta n}_i 
        =
        0
        .
    \end{aligned}
    \label{appx MHD: eqn33}
\end{equation}
To get \eqref{MHD:eqn10}, we only keep terms up to \(\mathcal{O}(k_y \rho_i)\).

\subsection{The role of incompressibility in MHD} \label{appx Ng MHD}

We shall demonstrate here how to recover the expressions in \citet{Ng05} by replacing \eqref{appx MHD: eqn7} with the incompressibility condition \(\nabla\cdot\boldsymbol{U} = 0\). In this case, equations \eqref{appx MHD: eqn24} and \eqref{appx MHD: eqn25} still holds, but \eqref{appx MHD: eqn26} simply reduces to only retaining the first two terms on the LHS:
\begin{equation}
    \begin{aligned}[b]
         \widehat{\delta u}_y = St\widehat{\delta u}_x 
         .
    \end{aligned}
    \label{appx Ng MHD:eqn1}
\end{equation}
Substituting into \eqref{appx MHD: eqn25} gives us:
\begin{equation}
    \begin{aligned}[b]
         \widehat{\delta u}_x = 
         -\left(\frac{\rmd \log n_i}{\rmd x} \right)^{-1} 
         \frac{\partial }{\partial t} \left(\frac{\widehat{\delta n}_i}{n_i} \right)
    \end{aligned}
    \label{appx Ng MHD:eqn2}
\end{equation}
substituting \eqref{appx Ng MHD:eqn1} and \eqref{appx Ng MHD:eqn2} into \eqref{appx MHD: eqn24} gives us
\begin{equation}
    \begin{aligned}[b]
         &
        (S^2t^2+1) \frac{\partial^2 \widehat{\delta n}_i}{\partial t^2}
        +
        2 S^2 t \frac{\partial \widehat{\delta n}_i}{\partial t} 
        =
         \frac{\omega_{gi}\omega_{*i}}{k_y^2 \rho_i^2/2} \widehat{\delta n}_i
        +
        \im\omega_{*i}\left(S^2 {t}^2 +1\right) \frac{\partial \widehat{\delta n}_i}{\partial t}
        +
        \im \omega_{*i} S^2 t\widehat{\delta n}_i
        ,
    \end{aligned}
    \label{appx Ng MHD:eqn3}
\end{equation}
which is equivalent to equation (16) in \citet{Ng05}. Because the incompressibility assumption is equivalent to dropping the second and third terms in \eqref{appx MHD: eqn7}, which in turn is equivalent to dropping the electron pressure term in Ohm's law \eqref{appx MHD: eqn3}, the resulting equation is much simpler than \eqref{appx MHD: eqn33}.

\section{Derivation of the gyrokinetic integral equation} \label{appx GK int}

We introduce the integrating factor \(e^{\im \omega_{gs} t}\) to equation \eqref{GK:eqn25}. This gives us
\begin{equation}
    \begin{aligned}
    \frac{\partial}{\partial t} 
	\left[
		e^{\im \omega_{gs} t}\hat{h}_{s} (t) 
	\right]
	=
	\frac{Z_s e F_{0s}}{T_s}
	e^{\im \omega_{gs} t}
	\left[
	\frac{\partial}{\partial t} 
	+
	\im (\omega_{gs}
	+
	\omega_{*s} )
	\right]
	J_0 \left(a_s(t) \right) \widehat{\delta\varphi} (t) 
    .
    \end{aligned}
    \label{appx GK int: eqn1}
\end{equation}
Integrating from \(0\) to \(t\), we then obtain
\begin{equation}
    \begin{aligned}
    e^{\im \omega_{gs} t}\hat{h}_{s} (t) 
	-
	\hat{h}_{s} (0) 
	=
	\frac{Z_s e F_{0s}}{T_s}
	\int_0^t \rmd t' \, 
	e^{\im \omega_{gs} t'}
	\left[
	\frac{\partial}{\partial t'} 
	+
	\im (\omega_{gs}
	+
	\omega_{*s} )
	\right]
	J_0 \left(a_s(t') \right) \widehat{\delta\varphi} (t')  
    ,
    \end{aligned}
    \label{appx GK int: eqn2}
\end{equation}
where \(t'\) is a dummy variable. The integral on the RHS of \eqref{appx GK int: eqn2} can be evaluated as
\begin{equation}
    \begin{aligned}
    &
	\int_0^t \rmd t' \, 
	e^{\im \omega_{gs} t'}
	\left[
	\frac{\partial}{\partial t'} 
	+
	\im (\omega_{gs}
	+
	\omega_{*s} )
	\right]
	J_0 \left(a_s(t') \right) \widehat{\delta\varphi} (t')  
    =
    -
    J_0 \left(a_s(0) \right) \widehat{\delta\varphi} (0)
    \\
    &+
    e^{\im \omega_{gs} t}  J_0 \left(a_s(t) \right) \widehat{\delta\varphi} (t)
    +
    \im \omega_{*s} \int_0^t \rmd t' \, e^{\im \omega_{gs} t'}  
    \left[ J_0 \left(a_s(t') \right) \widehat{\delta\varphi} (t')  \right]
    .
    \end{aligned}
    \label{appx GK int: eqn3}
\end{equation}
Substituting this back into \eqref{appx GK int: eqn2}, we obtain
\begin{equation}
    \begin{aligned}
        \hat{h}_s(t) 
         &=   
        \hat{h}_s(0) e^{-\im \omega_{gs} t}
        - 
        \frac{Z_s e F_{0s}}{T_s} e^{-\im \omega_{gs} t}J_0(a_s(0)) \widehat{\delta\varphi}(0)
        +
        \frac{Z_s e F_{0s}}{T_s} J_0(a_s(t)) \widehat{\delta\varphi}(t)
        \\
        & \hspace{0.5cm} +
        \frac{Z_s e F_{0s}}{T_s} \im \omega_{*s} 
        \int_0^{t} \rmd \Delta t \, e^{-\im \omega_{gs} \Delta t} J_0(a_s(t')) \widehat{\delta\varphi}(t - \Delta t)
        ,
    \end{aligned}
    \label{appx GK int: eqn4}
\end{equation}
where we have introduced another dummy variable \(\Delta t = t - t'\). 

\par 
Plugging \eqref{appx GK int: eqn4} into \eqref{GK:eqn26}, we obtain
\begin{equation}
    \begin{aligned}
        &\sum_s \frac{Z_s^2 e^2 n_s}{T_s} \widehat{\delta\varphi} (t) 
        =
        \sum_s Z_s e \int \rmd^3\boldsymbol{w} 
        J_0 \left(a_s(t) \right) e^{-\im \omega_{gs} t}
        \left[
            \hat{h}_s(0) 
            - 
            \frac{Z_s e F_{0s}}{T_s} J_0(a_s(0)) \widehat{\delta\varphi}(0)
        \right]
        \\
        &+\sum_s \frac{Z_s^2 e^2 n_s}{T_s} \frac{1}{\pi^{3/2} v_{\text{th},s}^3}
        \Bigg\{
        \widehat{\delta\varphi} (t) \int \rmd^3\boldsymbol{w}  \,
        e^{-w^2/v_{\text{th},s}^2}
        \Big[
        	J_0 \left(a_s(t) \right) 
        \Big]^2
        \\
        &+
        \im \omega_{*s} 
        \int_0^t \rmd \Delta t \, e^{-\im \omega_{gs} \Delta t}  
        \widehat{\delta\varphi} (t - \Delta t)
        \int \rmd^3\boldsymbol{w} 
        J_0 \left( a_s(t)\right)J_0 \left(a_s(t') \right) e^{-w^2/v_{\text{th},s}^2}
        \Bigg\}
        .
    \end{aligned}
    \label{appx GK int: eqn5}
\end{equation}
The velocity integrals of the second and third terms on the RHS can be evaluated using the identity \eqref{GK:eqn11}, resulting in
\begin{equation}
    \begin{aligned}[b]
    &\sum_s \frac{Z_s^2 e^2 n_s}{T_s} \widehat{\delta\varphi} (t)
    =
     \sum_s Z_s e \int \rmd^3\boldsymbol{w} 
        J_0 \left(a_s(t) \right) e^{-\im \omega_{gs} t}
        \left[
            \hat{h}_s(0) 
            - 
            \frac{Z_s e F_{0s}}{T_s} J_0(a_s(0)) \widehat{\delta\varphi}(0)
        \right]
    \\
    &+ 
    \sum_s \frac{Z_s^2 e^2 n_s}{T_s} 
    \left[ 
    \widehat{\delta\varphi} (t) 
    \Gamma_0(\lambda_s, \lambda_s)
    +
    \im \omega_{*s}  
    \int_0^t \rmd \Delta t \, e^{-\im \omega_{gs} \Delta t} 
    \Gamma_0(\lambda_s, \lambda_s')
    \widehat{\delta\varphi} (t - \Delta t)
    \right]
    ,
    \end{aligned}
    \label{appx GK int: eqn6}
\end{equation}
where we have used the definitions of \(\Gamma_0\) and \(\lambda_s\) in \eqref{GK:eqn28} and \eqref{GK:eqn29} respectively. Equation \eqref{appx GK int: eqn6} is equivalent to \eqref{GK:eqn27}.

\section{Derivation of the gyrokinetic perturbed ion density equation} \label{appx GK density}

\subsection{From the integral equation}

From the quasineutrality equation \eqref{GK:eqn26}, the perturbed ion density can be defined to be (\(Z_i = + 1\) for a hydrogen plasma):
\begin{equation}
    \begin{aligned}
    	\widehat{\delta n}_i 
    	&=
    	-
    	\frac{e n_i}{T_i} \widehat{\delta \varphi}
    	+
    	\int \rmd^3\boldsymbol{w} \, J_0(a_i(t)) \hat{h}_i 
        .
    \end{aligned}
    \label{appx GK density:eqn1}
\end{equation}
As we have done in the derivation of the gyrokinetic integral equation in \autoref{appx GK int}, we need to substitute the equation for \(\hat{h}_i\) in \eqref{appx GK int: eqn4} into \eqref{appx GK density:eqn1}. However, the resulting velocity integral is exactly the same as the one calculated in \eqref{appx GK int: eqn5}. The result is:
\begin{equation}
    \begin{aligned}
	\widehat{\delta n}_i 
	&=
	\frac{ e n_i}{T_i} 
    \left[
        \Gamma_0(\lambda_i, \lambda_i) - 1
    \right]
    \widehat{\delta \varphi}
	+
	\kappa n_i e^{-\im \omega_{gi} t}  e^{-\lambda_i(t)/2} 
    \\
    &\hspace{0.5cm}  + 
     \frac{ e n_i}{T_i} 
    \im \omega_{*i}  
    \int_0^t d\Delta t \, e^{-\im \omega_{gi} \Delta t} 
    \Gamma_0(\lambda_i, \lambda_i')
    \widehat{\delta\varphi} (t - \Delta t)
    .
    \end{aligned}
    \label{appx GK density:eqn2}
\end{equation}
At \(t = 0\), we have:
\begin{equation}
    \begin{aligned}
    	\widehat{\delta n}_i(0)
    	=
    	n_i 
    	\left\{
    		 \kappa e^{-\lambda_{i0}/2}
    		 +
    		 \frac{e }{T_i}
    		 \left[
    			 \Gamma_0(\lambda_{i0}, \lambda_{i0}) - 1 
    		 \right]\widehat{\delta\varphi} (0) 
    	\right\}
        .
    \end{aligned}
    \label{appx GK density:eqn3}
\end{equation}
Dividing \eqref{appx GK density:eqn2} by \eqref{appx GK density:eqn3}, using the definition of \(\widehat{\delta\varphi}(0)\) in \eqref{GK:eqn36}, undoing the change of variables \(\Delta t = t - t'\), and using the GX normalizations ultimately gives \eqref{numerical:eqn7}.

\subsection{From GX}

Using the GX normalizations, equation \eqref{appx GK density:eqn1} becomes:
\begin{equation}
    \begin{aligned}
    	\widetilde{\delta n}_i 
    	=
    	-
    	\widetilde{\delta \varphi} 
    	+
    	    \int \rmd^3 \boldsymbol{w} \, J_0(a_i(t)) \tilde{h}_i 
        .
    \end{aligned}
    \label{appx GK density:eqn4}
\end{equation}
The first and second quantities can be computed from the \texttt{Phi} and \texttt{ParticleDensity} diagnostic in GX. An important caveat is that the output of these quantities are in Fourier space and in the lab frame; however, the quantities that we are interested in are in the shearing frame. Fortunately, this is of no issue to us as it is straightforward to connect the shearing frame Fourier amplitudes with that of the lab frame.

\par 
To see this, we note that the shearing box transformation (see \S\ref{MHD SECT: shear box}) ensures that \(\bm{k}\cdot\bm{r} = \bm{k}'\cdot\bm{r}'\). Using \(\delta\varphi\) as an example, the Fourier transform in the shearing frame \eqref{GK:eqn22} can be more generally written as:
\begin{equation}
    \begin{aligned}
        \delta\varphi
        &=
        \sum_{k_y'} \sum_{k_x'}
         \widehat{\delta\varphi}(t', k_x', k_y') \, e^{\im k_x' x'}e^{ \im k_y' y'}
        =
         \sum_{k_y'} \sum_{k_x'}
         \widehat{\delta\varphi}(t', k_x', k_y') \, e^{\im k_x x} e^{ \im k_y y}
         .
    \end{aligned}
    \label{appx GK density:eqn5}
\end{equation}
On the other hand, in the lab frame, we also must have:
\begin{equation}
    \begin{aligned}
        \delta\varphi
        &=
        \sum_{k_y} \sum_{k_x}
         \widehat{\delta \psi}(t', k_x, k_y) \, e^{\im k_x x} e^{ \im k_y y}
         .
    \end{aligned}
    \label{appx GK density:eqn6}
\end{equation}
We are only concerned with \(k_x' = 0\) and a single \(k_y' = k_y\), so equating \eqref{appx GK density:eqn5} with \eqref{appx GK density:eqn6} gives:
\begin{equation}
    \begin{aligned}
         \widehat{\delta\varphi}(t', 0, k_y') 
        =
        \sum_{k_x}
        \widehat{\delta \psi}(t', k_x', k_y')
        \, e^{\im k_x x} 
         ,
    \end{aligned}
    \label{appx GK density:eqn7}
\end{equation}
where we have used the fact that from \eqref{MHD:eqn9}, we have \(k_x = - St' k_y'\) and \(k_y = k_y'\). Since we are interested in the volume average of the quantity \eqref{appx GK density:eqn6}, we have:
\begin{equation}
    \begin{aligned}
        \frac{1}{L_x} \int_0^{L_x} \rmd x \, 
         \widehat{\delta\varphi}(t', 0, k_y') 
        =
        \sum_{k_x}
        \left\lvert 
        \widehat{\delta \psi}(t', k_x', k_y')
        \right\rvert^2  
        ,
    \end{aligned}
    \label{appx GK density:eqn8}
\end{equation}
where we have invoked Parseval's theorem. It is the components \(\widehat{\delta \psi}(t', k_x', k_y')\) that is computed by GX (under the appropriate normalizations).

\par 
By this same logic, since \eqref{appx GK density:eqn4} is linear, we can add the complex values of the two diagnostics together on the RHS and then perform the summation as in the RHS of \eqref{appx GK density:eqn8}.

\section{Initial conditions for the finite shear MHD equation} \label{appx MHD init cond}

The finite shear MHD equation \eqref{numerical:eqn3} requires us to supplement the initial values of \(\widetilde{\delta n}_i(\tilde{t}) / \widetilde{\delta n}_i(0)\) and its derivative. While the former is clearly 1, the latter needs to  be obtained by differentiating \eqref{appx GK density:eqn1}:
\begin{equation}
    \begin{aligned}
	\left. \frac{\partial}{\partial t} \widehat{\delta n}_i \right\rvert_{t = 0}
	&=
	-
	\frac{e n_i}{T_i} 
	\left. \frac{\partial}{\partial t} \widehat{\delta \varphi} \right\rvert_{t = 0}
	+
	\int \rmd^3\boldsymbol{w} \, 
	\left. \frac{\partial}{\partial t} \left[ J_0(a_i(t)) \hat{h}_i  \right] \right\rvert_{t = 0}
    .
    \end{aligned}  
    \label{appx MHD init cond:eqn1}
\end{equation}

\par 
Let us first handle the second term on the RHS of \eqref{appx MHD init cond:eqn1}. Using the fact that \(J_0'(x) = - J_1(x)\), we can differentiate \eqref{appx GK int: eqn4} to obtain:
\begin{equation}
    \begin{aligned}
	\left. \frac{\partial}{\partial t} \hat{h}_i \right\rvert_{t=0}
	&=
    - \im\omega_{gi} \hat{h}_i(0)  
    +
    \frac{ e F_{0i}}{T_i}
    \im (\omega_{gi}+\omega_{*i} )
    J_0 \left(a_i(0) \right) \widehat{\delta\varphi} (0) 
     \\
    &\hspace{0.5cm}
    +
    \frac{ e F_{0i}}{T_i}
    J_0(a_i(0))
    \left. 
    \frac{\partial \widehat{\delta\varphi} (t) }{\partial t} 
    \right\rvert_{t = 0}
    -
    \frac{ e F_{0i}}{T_i}
    \widehat{\delta\varphi} (0)   \left. \frac{\partial a_i(t)}{\partial t} \right\rvert_{t =0} J_1(a_i(0))
    .
    \end{aligned}
    \label{appx MHD init cond:eqn2}
\end{equation}
Using our initial condition \eqref{GK:eqn32} for the gyrokinetic integral equation and the definitions of \(\lambda_i\) and \(a_i\) in \eqref{GK:eqn29}, we obtain:
\begin{equation}
    \begin{aligned}
	\left. \frac{\partial}{\partial t} \left[ J_0(a_i(t)) \hat{h}_i  \right] \right\rvert_{t = 0}
	&=
    - \im \kappa \omega_{gi} J_0(a_i(0)) F_{0i}  
    +
    \frac{ e F_{0i}}{T_i}
    \left[ J_0(a_i(0)) \right]^2
    \left. 
    \frac{\partial \widehat{\delta\varphi} (t) }{\partial t} 
    \right\rvert_{t = 0}
    \\
    & \hspace{0.5cm}+
    \im \omega_{*i}  \frac{ e \widehat{\delta\varphi} (0) }{T_i} 
    \left[ J_0(a_i(0)) \right]^2 F_{0i}  
    .
    \end{aligned}
    \label{appx MHD init cond:eqn3}
\end{equation}
The velocity integration then gives:
\begin{equation}
    \begin{aligned}
    	\int \rmd^3 \boldsymbol{w}
    	\left. \frac{\partial}{\partial t} \left[ J_0(a_i(t)) \hat{h}_i  \right] \right\rvert_{t = 0}
    	&=
        \frac{ e n_i}{T_i}
        \left[
        \left. 
        \frac{\partial \widehat{\delta\varphi} (t) }{\partial t} 
        \right\rvert_{t = 0}
        +
        \im \omega_{*i} \widehat{\delta\varphi} (0) 
        \right]
        \Gamma_0 \left(\lambda_{i0}, \lambda_{i0} \right)
        \\
        &\hspace{0.5cm} -
        \im \kappa \omega_{gi} n_i e^{-\lambda_{i0}/2}
        .
    \end{aligned}
    \label{appx MHD init cond:eqn4}
\end{equation}

\par 
The first term on the RHS of \eqref{appx MHD init cond:eqn1} can be calculated directly from the integral equation \eqref{GK:eqn37} where it suffices to take the small mass ratio approximation \(m_e/m_i \ll 1\) as it is an MHD calculation:
\begin{equation}
    \begin{aligned}
	&
	\widehat{\delta\varphi}(t)	
	\approx
	\frac{\kappa T_i}{e} 
	\frac{
		e^{-\lambda_i/2} 
		e^{-\im\omega_{gi}t}
		-
		1
	}{
		1 - \Gamma_0(\lambda_i, \lambda_i)
	} 
	+
	 \im \omega_{*i}
	 \int_0^t \rmd t'\,
	 \frac{
		 e^{- \im \omega_{gi} (t-t')} \Gamma_0(\lambda_i, \lambda_i') 
		  -
		 1
	}{
		1 - \Gamma_0(\lambda_i, \lambda_i)
	}
	\widehat{\delta\varphi}(t')
	.
    \end{aligned}
    \label{appx MHD init cond:eqn5}
\end{equation}
By noting that \(I_0'(x) = I_1(x)\), we have:
\begin{align}
    &
    \begin{aligned}[b]
	\frac{\partial}{\partial t} \Gamma_0(\lambda_i, \lambda_i')
	=
	\frac{1}{2}\Gamma_0(\lambda_i, \lambda_i')
	\left[
		\frac{1}{\sqrt{\lambda_i \lambda_i'} }
		\frac{I_1 \left(\sqrt{\lambda_i \lambda_i'} \right)}{I_0 \left(\sqrt{\lambda_i \lambda_i'} \right)}
		\lambda_i'
		-
		1
	\right]
	\frac{\partial \lambda_i}{\partial t}
    ,
    \end{aligned}
    \label{appx MHD init cond:eqn6}
    \\
    &
    \frac{\partial}{\partial t} \Gamma_0(\lambda_i, \lambda_i)
	=
	\frac{\partial \lambda_i}{\partial t}
	\left[
		\frac{ I_1(\lambda_i)}{I_0(\lambda_i)} - 1
	\right]\Gamma_0(\lambda_i, \lambda_i)
    .
    \label{appx MHD init cond:eqn7}
\end{align}
This allows us to differentiate \eqref{appx MHD init cond:eqn5} and obtain:
\begin{equation}
    \begin{aligned}
	&\frac{\partial}{\partial t}
	\widehat{\delta\varphi}(t)	
    \\
	&=
	\frac{\kappa T_i/e}{1 - \Gamma_0(\lambda_i, \lambda_i)}
	\left[
		-
		\left(
			\frac{1}{2}
			\frac{\partial \lambda_i}{\partial t}
			+
			\im\omega_{gi}
		\right)
		e^{-\lambda_i/2} 
		e^{-\im\omega_{gi}t}
		+
		\frac{e^{-\lambda_i/2} 
				e^{-\im\omega_{gi}t}
				-1
				}{1 - \Gamma_0(\lambda_i, \lambda_i) } 
		\frac{\partial \Gamma_0(\lambda_i, \lambda_i)}{\partial t}
	\right]
	\\
	&\hspace{0.5cm}
	+ 
	\im \omega_{*i}
	\left\{
		\int_0^t \rmd t' \, 
		\left[ 
			e^{- \im \omega_{gi} (t-t')} \Gamma_0(\lambda_i, \lambda_i') 
			-1
		\right]
		\widehat{\delta\varphi}(t')
	\right\}
	\frac{\partial}{\partial t}
	\frac{1}{1 - \Gamma_0(\lambda_i, \lambda_i)}
	\\
	&\hspace{0.5cm}
	-
	\im \omega_{*i}
	\widehat{\delta\varphi}(t)
	+
	\frac{\im \omega_{*i}}{1 - \Gamma_0(\lambda_i, \lambda_i)}
	\int_0^t \rmd t' \, 
	\frac{\partial}{\partial t}
	\left\{
	\left[ 
		e^{- \im \omega_{gi} (t-t')} \Gamma_0(\lambda_i, \lambda_i') 
		-1
	\right]
	\widehat{\delta\varphi}(t')
	\right\}
    ,
    \end{aligned}
    \label{appx MHD init cond:eqn8}
\end{equation}
and hence:
\begin{equation}
    \begin{aligned}
	\left. \frac{\partial}{\partial t}
	\widehat{\delta\varphi}(t)	\right\rvert_{t=0}
	&=
	-
	\frac{\kappa T_i}{e}
	\frac{\im\omega_{gi}}{1 - \Gamma_0(\lambda_{i0}, \lambda_{i0})}
		e^{-\lambda_{i0}/2} 
	-
	\im \omega_{*i}
	\widehat{\delta\varphi}(0)
    .
    \end{aligned}
    \label{appx MHD init cond:eqn9}
\end{equation}

\par
Substituting the results \eqref{appx MHD init cond:eqn4} and \eqref{appx MHD init cond:eqn9} into \eqref{appx MHD init cond:eqn1}, we obtain:
\begin{equation}
    \begin{aligned}
        \left. \frac{\partial}{\partial t} \frac{\widehat{\delta n}_i}{n_i} \right\rvert_{t = 0}
    	=
    	\im \kappa \omega_{*i}
    	\frac{
    		e^{-\lambda_{i0}/2} - 1
    	}{
    		1 - \Gamma_0(\lambda_{i0}, \lambda_{i0})
    	}
        ,
    \end{aligned}
    \label{appx MHD init cond:eqn10}
\end{equation}
where we have used and taken the small mass ratio \(m_e/m_i \ll 1\) of \(\widehat{\delta\varphi}(0)\) given in \eqref{GK:eqn36}. In addition, this also allows us to obtain \(\widehat{\delta n}_i(0) \approx \kappa n_i\) from \eqref{appx GK density:eqn3}, for which we can then rewrite \eqref{appx MHD init cond:eqn10} as:
\begin{equation}
    \begin{aligned}
         \left. \frac{\partial}{\partial t} \frac{\widehat{\delta n}_i}{\widehat{\delta n}_i(0)} \right\rvert_{t = 0}
	=
	\im  \omega_{*i}
	\frac{
		e^{-\lambda_{i0}/2} - 1
	}{
		1 - \Gamma_0(\lambda_{i0}, \lambda_{i0})
	} 
    .
    \end{aligned}
    \label{appx MHD init cond:eqn11}
\end{equation}
Taking the long-wavelength limit \(\lambda_{i0}\ll1\) and using the GX normalizations, we finally arrive at:
\begin{equation}
    \begin{aligned}
         \left. \frac{\partial}{\partial \tilde{t}} \frac{\widetilde{\delta n}_i}{\widetilde{\delta n}_i(0)} \right\rvert_{t = 0}
        \approx
        - \im \frac{\tilde{\omega}_{*i}}{2} 
        .
    \end{aligned}
    \label{appx MHD init cond:eqn12}
\end{equation}
It is this expression we shall use to compute our initial derivative for the finite shear MHD equation.

\section{Low order GK expansions} \label{appx GK low order}

\subsection{Electrostatic potential equation}

Taking the \(m_e/m_i = \tilde{m}_e \ll 1\) limit of the integral equation for \(\widetilde{\delta\varphi}\) \eqref{numerical:eqn4} gives:
\begin{equation}
    \begin{aligned}
    	\frac{\widetilde{\delta\varphi}(\widetilde{t})}{\widetilde{\delta\varphi}(0)}
    	&=
    	\frac{
    		e^{-\lambda_i/2} e^{-\text{i}\widetilde{\omega}_{gi} \widetilde{t}}
    		-
    		1
    	}{
    		e^{-\lambda_{i0}/2} - 1
    	}
    	\frac{
    		1 - \Gamma_0(\lambda_{i0}, \lambda_{i0})
    	}{
    		1 - \Gamma_0(\lambda_i, \lambda_i)
    	}
    \\
	   &\hspace{0.5cm} +
	   \text{i} \widetilde{\omega}_{*i}
	   \int_0^{\tilde{t}} \rmd \tilde{t}'\, 
	   \frac{
		e^{- \text{i} \widetilde{\omega}_{gi} (\widetilde{t}-\widetilde{t}')} \Gamma_0(\lambda_i, \lambda_i') 
		-
		1
	       }{
		1 - \Gamma_0(\lambda_i, \lambda_i)
	       }
    \frac{\widetilde{\delta\varphi}(\widetilde{t}')}{\widetilde{\delta\varphi}(0)}
    .
    \end{aligned}
    \label{appx GK low order:eqn1}
\end{equation}
Next, by assuming \(\tilde{k}_y \ll 1\) and hence \(\lambda_i,\lambda_i' \sim \tilde{k}_y^2 \ll 1\), we have:
\begin{align}
    &
    \Gamma_0(\lambda_i, \lambda_i) \approx
    1 - \lambda_i + \frac{3}{4} \lambda_i^2 
    ,
    \label{appx GK low order:eqn2}
    \\
    &
    \Gamma_0(\lambda_i, \lambda_i')
    \approx
    1 - \frac{\lambda_i + \lambda_i'}{2}
	+ \frac{\lambda_i^2 + {\lambda_i'}^2}{8}  + \frac{\lambda_i \lambda_i'}{2}
    .
    \label{appx GK low order:eqn3}
\end{align} 
Noting that \(\tilde{\omega}_{gi}\sim\tilde{\omega}_{*i} \sim \tilde{k}_y\), equation \eqref{appx GK low order:eqn1} to \(\mathcal{O}(\tilde{k}_y)\) is given by:
\begin{equation}
    \begin{aligned}
	\frac{\widetilde{\delta\varphi}(\widetilde{t})}{\widetilde{\delta\varphi}(0)}
	&=
	-
	\frac{2}{\lambda_i}
	\left(
		- \text{i}\widetilde{\omega}_{gi}\tilde{t} -\frac{\widetilde{\omega}_{gi}^2\tilde{t}^2}{2}
		+ \frac{\text{i}\widetilde{\omega}_{gi}^3\tilde{t}^3}{6}
	\right)
	-
	\frac{5}{2}
	\left(
		1 - \text{i}\widetilde{\omega}_{gi}\tilde{t} 
	\right)
	+
    	 \frac{3}{2} 
    	+
    	\frac{1}{2}\frac{\lambda_{i0}}{\lambda_i}
    	\text{i}\widetilde{\omega}_{gi}\tilde{t} 
    	\\
    	&\hspace{0.5cm} +
    	\text{i} \widetilde{\omega}_{*i}
    	\int_0^{\tilde{t}} \rmd \tilde{t}'\, 
    	\frac{\widetilde{\delta\varphi}(\widetilde{t}')}{\widetilde{\delta\varphi}(0)}
    	\left[
    		-\frac{1}{2} 
    		-
    		\frac{\lambda_i'}{2 \lambda_i}
    		- 
    		\frac{\text{i} \widetilde{\omega}_{gi}}{\lambda_i} (\widetilde{t}-\widetilde{t}')
    		- 
    		\frac{1}{2\lambda_i}\widetilde{\omega}_{gi}^2 (\widetilde{t}-\widetilde{t}')^2
    	\right]
        .
    \end{aligned}
    \label{appx GK low order:eqn4}
\end{equation}

\subsection{Perturbed density equation}

We now need to repeat the procedure for the perturbed density equation \eqref{numerical:eqn7}. The result is:
\begin{equation}
    \begin{aligned}
	   \frac{\widetilde{\delta n}_i }{\widetilde{\delta n}_i (0)}
	&\approx
	1  - \text{i}\tilde{\omega}_{gi} \tilde{t}  
    -\frac{\text{i} \tilde{\omega}_{*i} }{2}
    \int_0^{\tilde{t}} \rmd \tilde{t}' \, 
	\frac{\widetilde{\delta\varphi} (\tilde{t}')}{\widetilde{\delta\varphi} (0)}
    .
    \end{aligned}
    \label{appx GK low order:eqn5}
\end{equation}
The combination of \eqref{appx GK low order:eqn4} and \eqref{appx GK low order:eqn5} allows us to obtain the grey line in \autoref{numerical:fig2}.

\bibliographystyle{jpp}

\bibliography{interchange_jpp}

\end{document}